\documentclass[twocolumn]{aastex63}

\usepackage{lineno}
\usepackage[utf8]{inputenc}
\usepackage{amsmath}
\usepackage{epstopdf}
\usepackage{gensymb}
\usepackage{natbib}
\usepackage{CJKutf8}

\shorttitle{PAH-Metallicity Relation}
\shortauthors{Whitcomb et al.}

\newcommand{\totpah}{$\Sigma$PAH}
\newcommand{\pahtir}{$\Sigma$PAH/TIR}
\newcommand{\qpah}{$q_{\rm {PAH}}$}

\newcommand{\elevteen}{PAH\,11.3/17\,\micron}
\newcommand{\sevelev}{PAH\,7.7/11.3\,\micron}
\newcommand{\sixsev}{PAH\,6.2/7.7\,\micron}
\newcommand{\sixteen}{PAH\,6.2/17\,\micron}

\newcommand{\rattot}{PAH/$\Sigma$PAH}
\newcommand{\sixtot}{PAH\,6.2/$\Sigma$PAH}
\newcommand{\sevtot}{PAH\,7.7/$\Sigma$PAH}
\newcommand{\elevtot}{PAH\,11.3/$\Sigma$PAH}
\newcommand{\teentot}{PAH\,17/$\Sigma$PAH}

\newcommand{\hnu}{$\langle h\nu\rangle$}

\newcommand{\carbons}{$N_{\rm C}$}
\newcommand{\aone}{$a_{\rm 01}$}

\newcommand{\twelog}{12~+~[O/H]}
\newcommand{\zmetal}{$Z$}
\newcommand{\zsolar}{$Z_{\odot}$}
\newcommand{\zcent}{$Z_{o}$}
\newcommand{\zthresh}{$Z_\mathrm{th}$}

\begin{document}

\title{The Metallicity Dependence of PAH Emission in Galaxies I: \\Insights from Deep Radial Spitzer Spectroscopy}

\author[0000-0003-2093-4452]{Cory M. Whitcomb}
\affil{Ritter Astrophysical Research Center, Department of Physics \& Astronomy, University of Toledo, Toledo, OH 43606, USA}

\author[0000-0003-1545-5078]{J.-D. T. Smith}
\affil{Ritter Astrophysical Research Center, Department of Physics \& Astronomy, University of Toledo, Toledo, OH 43606, USA}

\author[0000-0002-4378-8534]{Karin Sandstrom}
\affil{Department of Astronomy \& Astrophysics, University of California, San Diego, 9500 Gilman Drive, La Jolla, CA 92093, USA}

\author[0000-0003-2952-5444]{Carl A. Starkey}
\affil{Ritter Astrophysical Research Center, Department of Physics \& Astronomy, University of Toledo, Toledo, OH 43606, USA}
\affil{The Ohio State University Wexner Medical Center, 460 W 12th Ave, Columbus, OH 43210, USA}

\author[0009-0001-6065-0414]{Grant P. Donnelly}
\affil{Ritter Astrophysical Research Center, Department of Physics \& Astronomy, University of Toledo, Toledo, OH 43606, USA}

\author[0000-0002-0846-936X]{Bruce T. Draine}
\affil{Department of Astrophysical Sciences, Princeton University, Princeton, NJ 08544, USA}

\author[0000-0003-0605-8732]{Evan D. Skillman}
\affiliation{University of Minnesota, Minnesota Institute for Astrophysics, School of Physics and Astronomy,\\ 116 Church Street, S.E., Minneapolis, MN 55455, USA}

\author[0000-0002-5782-9093]{Daniel~A.~Dale}
\affiliation{Department of Physics and Astronomy, University of Wyoming, Laramie, WY 82071, USA}

\author[0000-0003-3498-2973]{Lee Armus}
\affil{IPAC, California Institute of Technology, 1200 E. California Boulevard, Pasadena, CA 91125, USA}

\author[0000-0001-7449-4638]{Brandon S. Hensley}
\affil{Jet Propulsion Laboratory, California Institute of Technology, 4800 Oak Grove Drive, Pasadena, CA, USA}

\author[0000-0001-8490-6632]{Thomas S.-Y. Lai\begin{CJK*}{UTF8}{bsmi} (賴劭愉)\end{CJK*}}
\affil{IPAC, California Institute of Technology, 1200 E. California Boulevard, Pasadena, CA 91125, USA}

\author[0000-0001-5448-1821]{Robert C. Kennicutt}
\affil{Steward Observatory, University of Arizona, Tucson, AZ 85719, USA}
\affil{George P. and Cynthia W. Mitchell Institute for Fundamental Physics \& Astronomy, Texas A\&M University, College Station, TX  77843, USA}

\correspondingauthor{Cory M. Whitcomb}
\email{coryw777@gmail.com}

\begin{abstract} 
We use deep Spitzer mid-infrared spectroscopic maps of radial strips across three nearby galaxies with well-studied metallicity gradients (M101, NGC~628, and NGC~2403) to explore the physical origins of the observed deficit of polycyclic aromatic hydrocarbons (PAHs) at sub-solar metallicity (i.e., the PAH-metallicity relation or PZR). These maps allow us to trace the evolution of all PAH features from 5--18\,\micron\ as metallicity decreases continuously from solar (\zsolar) to 0.2\,\zsolar. The total PAH to dust luminosity ratio remains relatively constant until reaching a threshold of $\sim\frac{2}{3}$\,\zsolar, below which it declines smoothly but rapidly. The PZR has been attributed to preferential destruction of the smallest grains in the hard radiation environments found at low metallicity.  In this scenario, a decrease in emission from the shortest wavelength PAH features is expected.  In contrast, we find a steep decline in long wavelength power below \zsolar, especially in the 17\,\micron\ feature, with the shorter wavelength PAH bands carrying an increasingly large fraction of power at low metallicity. We use newly developed grain models to reproduce the observed PZR trends, including these variations in fractional PAH feature strengths. The model that best reproduces the data employs an evolving grain size distribution that shifts to smaller sizes as metallicity declines. We interpret this as a result of inhibited grain growth at low metallicity, suggesting continuous replenishment in the interstellar medium is the dominant process shaping the PAH grain population in galaxies.

\end{abstract}


\section{Introduction}\label{sec:intro}
The smallest carbonaceous dust grains in the interstellar medium emit bright, broad vibrational features in the mid-infrared. These dust grains are commonly believed to be polycyclic aromatic hydrocarbons (PAHs) composed of dozens to thousands of aromatically-bonded carbon atoms \citep{tiel08, li20}. The emission from these PAH features alone contributes over 10\% of the total infrared luminosity of star-forming galaxies \citep{jd07}. 

Many previous studies across a wide variety of interstellar conditions have found deficits in the fraction of PAH luminosity relative to total dust continuum power at sub-solar metallicity, typically interpreted as a marked change in the abundance of dust in the form of PAHs \citep[e.g.][]{engel05,madden06,jackson06,draine07sings,jd07, gordon08,munoz09,haynes10,sandstrom12,lai20,aniano20,whitcomb20,zang22, shivaei24}. Little is known about the origins of this deficit. In the era of JWST it is crucial that we understand this PAH-metallicity relation (PZR) in order to accurately interpret PAH emission from local highly-resolved galaxies to systems at high redshift where metallicities for a given stellar mass may be significantly lower.

A common explanation for the PZR is that smaller PAHs are destroyed by high energy photons in low metallicity regions \citep{madden06, gordon08, egorov23}. At low metallicity, young stars are hotter and less dust is present to attenuate their hard photons \citep{massey05}. The resulting interstellar radiation fields (ISRFs) have a higher ratio of UV to optical photons, increasing the stochastic heating rate and peak temperature of PAHs. In general the heat capacity of dust grains is proportional to their volume, so the smallest PAH molecules may begin to dissociate when they absorb photons of sufficiently high energy \citep{guha89, lepage03, duley09, D21}.

In this paper we investigate the PZR using large, deep, Spitzer spectral maps (5--18\,\micron) in radial strips across three nearby galaxies --- M101, NGC~628, and NGC~2403 --- with well-studied abundance gradients obtained using auroral-line based measurements \citep{kenn03, croxall16, berg13, berg15, chaosVI}. These mid-infrared (MIR) spectral mapping datasets are some of the deepest ever observed with the Spitzer-InfraRed Spectrograph (IRS), and extend up to $\sim$15~kpc from galaxy center, enabling variations in the full PAH emission spectrum to be tracked, aside from the weak 3.3\,\micron\ band. By studying how the PAH spectrum changes in strength and shape as a function of radius within each galaxy, we can isolate the variations that result primarily from the radial metallicity gradient. For the first time, we model the recovered PZR trends using detailed physical models of PAH emission in the context of varying incident radiation fields and underlying grain size distributions (GSDs).

This paper is organized as follows. In \S\,\ref{sec:background} we summarize the key physical processes and potential explanations for the observed PZR trends. In \S\,\ref{sec:data} we describe our data, region definition criteria, and the resulting suite of data products.  In \S\,\ref{sec:obsresults} we describe details of the observed PZR trends. In \S\,\ref{sec:dvresults} we introduce several physically-motivated model scenarios and test them against the observations.  In \S\,\ref{sec:disc} we discuss the physical implications of our results and compare them with previous studies. In \S\,\ref{sec:concl} we outline our most significant conclusions.

\begin{figure*}
\centering
\includegraphics[width=\linewidth]{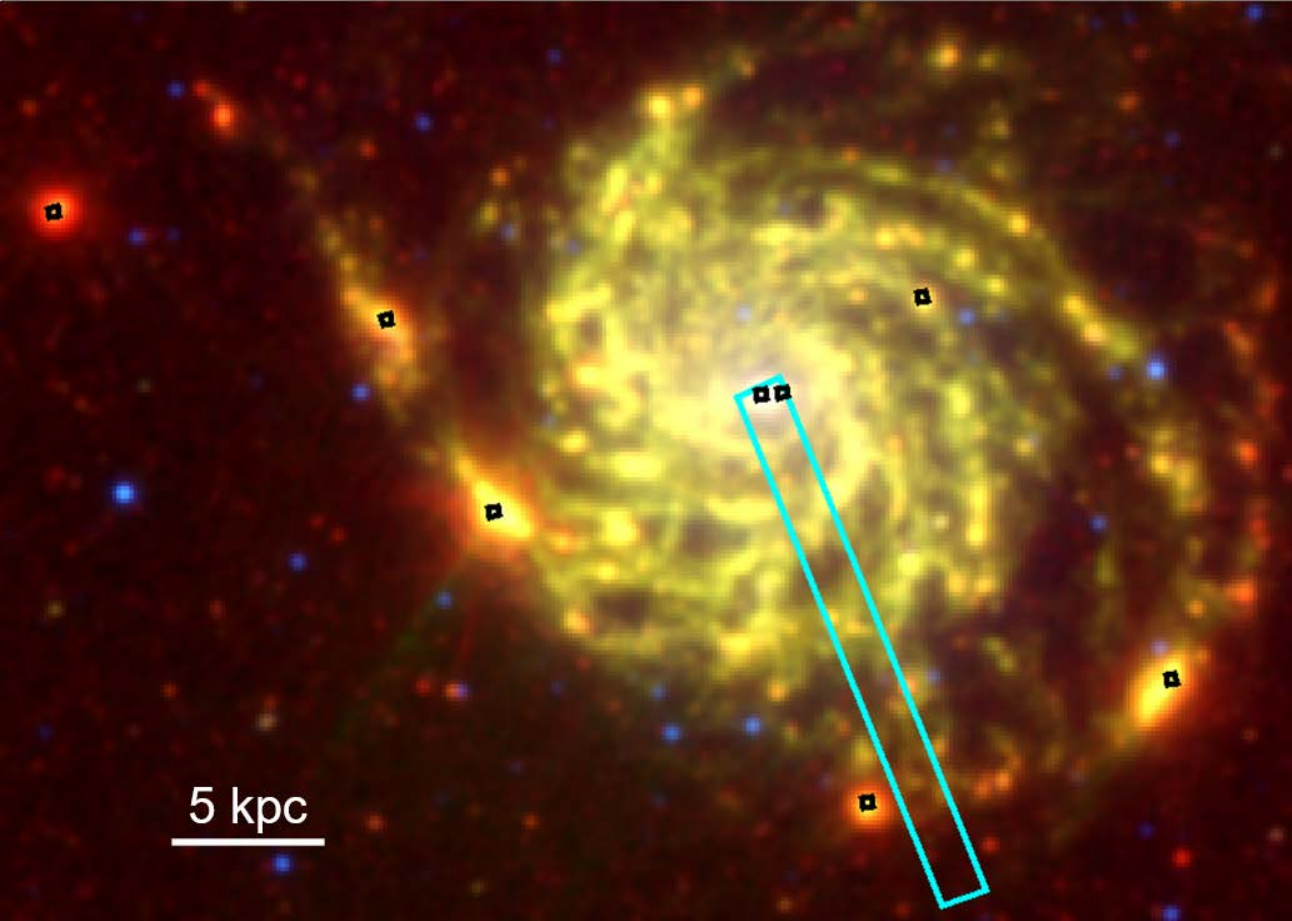}
\includegraphics[width=\linewidth]{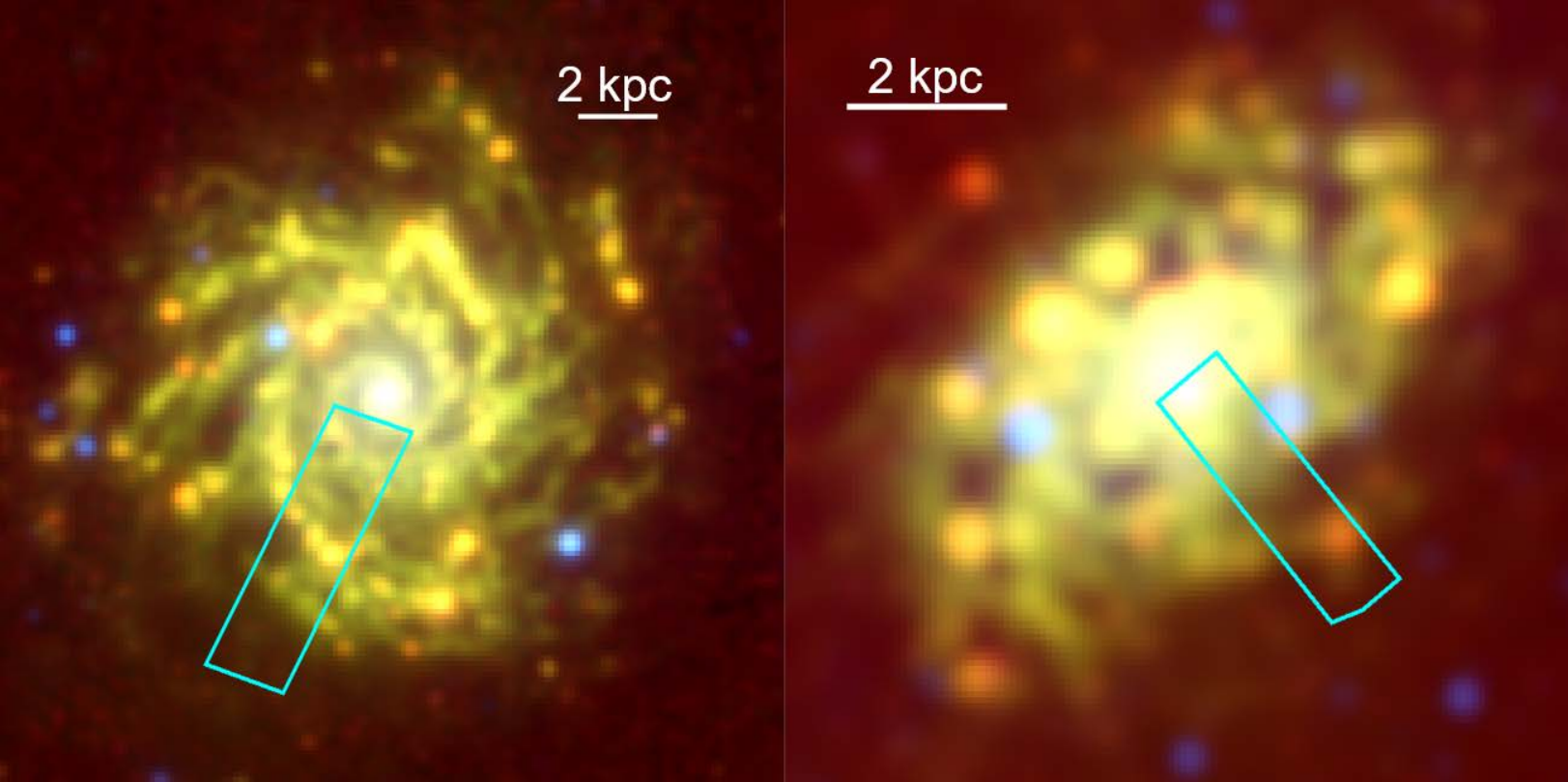}
\caption{\textbf{Three-color images for M101 (top), NGC~628 (left), and NGC~2403 (right).} Red is MIPS 24\,\micron, green is IRAC 8\,\micron, and blue is IRAC 3.6\,\micron. All images are convolved to match PACS 160\,\micron\ resolution, except those for NGC~2403 which are convolved to a 25\arcsec\ Gaussian. The footprint of the radial spectroscopic maps studied in this work are shown in cyan, and the eight additional apertures for the \textsc{Hii} region maps in M101 are shown as black squares.}
\label{figure:prettyrgb}
\end{figure*}

\section{Background and Definitions}\label{sec:background}

As the gas-phase metal abundance of the interstellar medium (ISM) drops, the dust-to-gas mass ratio decreases towards the background level set by stellar production \citep{remy14}, or lower in galaxies where grain destruction is more efficient than grain growth \citep{aniano20}. The resulting lack of dust opacity increases the average hardness \hnu\ and intensity $U$ of the starlight incident on dust grains.  Just as important as the heating environment, the distribution of grain sizes plays a key role in the emerging PAH spectrum.

Leading models suggest the distribution of grain sizes results from a complex balance between many opposing forces: injection in carbon-rich AGB winds, destruction in supernova shocks and ionized gas, fragmentation of larger grains into smaller units, coagulation and accretion building grain mass in dense clouds, etc. \citep{dwek80, zhukovska08, draine09, micel10a, galliano18, choban22, desika23, choban24}. Many of these processes are sensitive to metal abundance. For example, the timescale of gas-phase accretion onto grain surfaces has a strong dependence on the free metal content of the ISM \citep{zhukovska08, zhukovska13}.

One of the earliest Spitzer results on PAHs revealed a striking difference in PAH emission between galaxies with solar metallicity and those below one-quarter solar: over this range, PAH emission declines rapidly \citep{engel05, ohall06, jd07, hunt10, remy15, shivaei17, chaste19, aniano20}. It remains unclear whether the PAH abundance fraction relative to the total dust mass -- \qpah\ -- drops suddenly or smoothly as metallicity decreases. Some studies find \qpah\ remains constant at about 4\%, then drops suddenly to about 1\% after a threshold at about 0.2\,-\,0.3 \zsolar\ \citep{draine07sings, chaste19, shim23}. Others find a continuous decrease in the PAH fraction as a function of metallicity \citep{galliano08, remy15, galliano18,  aniano20}.

It has been common to study the PZR using integrated galaxy measurements from galaxies across a wide range of metallicity \citep{engel05, draine07sings, jd07, engel08, hunt10, shim23}. This method introduces complications because the radiation environments and star formation histories vary between galaxies. Resolved studies within individual galaxies have also found similar deficits in relative PAH to dust luminosity as metallicity decreases \citep{lebouteiller11, khramtsova14, chaste23}. Within galaxies the main properties that vary with radius on scales larger than 100 pc are the stellar density, stellar age, and gas-phase metal abundance. As the density of stars decreases at larger radii, the average intensity $U$ of the ISRF decreases as well. The average age of stars and the ISM density also decrease at larger radii \citep{garcia17, dale20}, increasing the average hardness \hnu\ of the ISRF.

The effect each of these properties of the radiation environment has on the PAH spectrum can be explicitly modeled with the new suite of theoretical spectra from \citet{D21} (hereafter D21).  The D21 models are based on the simplifying assumption that PAHs can be idealized as nearly spherical grains of pure graphite characterized by a single size parameter and a binary ionization state. This is sufficient to produce MIR emission features that match observations of galaxies. Other models use spectra of specific PAH molecules from quantum-chemical calculations, however this is currently only feasible for a small fraction of the PAH size distribution \citep{rigopoulou21}. Since we are interested in exploring the spectrum produced by the full range of PAH sizes exposed to various radiation environments, the D21 models provide an excellent framework for comparison to the observations.

Resolved PZR studies probing the centers of individual star-forming spiral galaxies using photometric tracers of PAH emission have been conducted with JWST imaging. These have well-constrained metallicity gradients in the radial direction, and the fraction \qpah\ is found to decrease relatively smoothly \citep{chaste23}. Most previous studies on the PZR only make use of photometric PAH tracers, primarily at 8\,\micron\ to study the strongest PAH feature at 7.7\,\micron. This feature is primarily emitted by ionized PAHs with $\sim100$~carbon atoms (\carbons), resulting in the common explanation that the smallest PAHs are being preferentially destroyed in low metallicity environments. The full PAH spectrum has contributions from both ionized and neutral grains with $\sim$20~--~10$^6$~\carbons. A full accounting of the properties of the small grain population requires spectroscopic access to all the main PAH emission bands. 

In addition to changes in the overall PAH luminosity, we can track several key variations in the PAH spectrum itself as metallicity decreases. Based on the theoretical work of \citet{lidraine01, draineli07, D21}, each PAH feature traces a specific regime of the grain size and ionization distributions of the local PAH population. In general, small PAHs emit more efficiently at shorter wavelengths and large PAHs emit more efficiently at longer wavelengths. Specifically, the approximate peaks are as follows \citep[][see Figure 6]{D21}: the 17\,\micron\ feature is emitted most efficiently by PAHs with $\sim2,000$~\carbons\ with no ionization preference, 11.3\,\micron\ by neutral PAHs with $\sim400$~\carbons, 7.7\,\micron\ by ionized PAHs with $\sim100$~\carbons, and 6.2\,\micron\ by ionized PAHs with $\sim70$~\carbons. The final major PAH feature at 3.3\,\micron\ is emitted at peak efficiency by neutral PAHs just above the minimum survivable PAH size adopted in D21 of 27~\carbons. Since this work uses Spitzer spectroscopy from 5--38\,\micron, we do not have 3.3\,\micron\ coverage. Instead we use the best fit models to predict the metallicity trends in this final PAH feature.

\citet{jd07} found the ratio of the 17 to 11.3\,\micron\ features drops significantly as metallicity declines. This was attributed to enhanced formation of grains larger than $\sim400$~\carbons\ at higher metallicities. \citet{sandstrom12} found a similar result: the 17\,\micron\ feature is significantly weaker, but still detected, in low metallicity regions of the Small Magellanic Cloud (SMC) compared to the higher metallicity galaxies of the Spitzer Infrared Nearby Galaxies Survey \citep[SINGS;][]{jd07}. They found the 7.7\,\micron\ feature is also significantly weaker, and the 6.2 and 11.3\,\micron\ features carry an increased fraction of the PAH power. This suggests PAHs in the low metallicity SMC are smaller and more neutral. \citet{hunt10} found the fractional PAH power in the 8.6\,\micron\ and 11.3\,\micron\ features is increased for low-metallicity blue compact dwarf galaxies compared to the SINGS galaxies, while the 6.2\,\micron\ and 7.7\,\micron\ features remain constant. They interpret this as evidence for an decrease in the PAH population smaller than 100~\carbons\ in low metallicity galaxies compared to high metallicity systems.

Studies of the PZR in galaxies have resulted in many different interpretations and several questions remain unanswered. Some attribute the PAH deficit to timing: a galaxy with low metallicity and a young stellar population simply has not had enough time to build up a significant reservoir of PAHs yet \citep{galliano08}. However, most agree that the relative PAH-to-dust luminosity drops as a result of (or correlated with) the decreased gas-phase metal abundance \citep{ohall06, draine07sings, gordon08, hunt10, shivaei17, aniano20, shivaei24}. 

\begin{deluxetable}{cccc}
\caption{Spitzer-IRS Observations\label{table: datainfo}}
\tablehead{
\colhead{\textbf{Target Name}} & 
\colhead{\textbf{RA (J2000)}} &
\colhead{\textbf{Dec (J2000)}} &
\colhead{\textbf{AORID}}
}
\startdata
\textbf{M101} \\
SLmap\textunderscore 01	&	14h03m01.90s &	+54d17m01.0s &	14800128	\\
SLmap\textunderscore 00	&	14h03m10.50s &	+54d20m10.8s &	14799872	\\
SLmap\textunderscore 02	&	14h02m53.30s &	+54d13m51.2s &	14800384	\\
SL BG                   &	14h03m26.50s &	+54d37m35.0s &	14800896	\\
LLmap                   &	14h03m01.90s &	+54d17m01.0s &	14799616	\\
LL BG                   &	14h03m26.50s &	+54d37m35.0s &	14800640	\\
\textbf{NGC 628} \\
SLmap\textunderscore 00	&	1h36m48.10s  &	+15d43m56.2s &	14802432	\\
SLmap\textunderscore 01	&	1h36m43.30s  &	+15d46m14.4s &	14802688	\\
SL BG					&	1h36m54.13s  &	+15d37m59.0s &	14803200	\\
LLmap                   &	1h36m45.70s  &	+15d45m05.3s &	14802944	\\
LL BG					&	1h36m54.13s  &	+15d37m59.0s &	14803456	\\
\textbf{NGC 2403} \\
SLmap\textunderscore 01	&	7h36m44.70s  &	+65d35m23.7s &	14801408\\
SLmap\textunderscore 00	&	7h36m28.90s  &	+65d33m24.5s &	14801152\\
SL BG					&	7h36m29.69s  &	+65d27m27.6s &	14802176	\\
LLmap                   &	7h36m36.80s  &	+65d34m24.1s &	14801664\\
LL BG					&	7h36m29.69s  &	+65d27m27.6s &	14801920\\
\enddata
\end{deluxetable}

Interpretation of the observed PZR trends is further complicated by the many unknowns about how small grains are produced and survive in the ISM. It is still unclear if the trends are a result of changes to the initial PAH GSD from stellar injection or changes in formation and processing pathways in the ISM. What the dominant sources of grains that emit as PAHs are and how their formation mechanisms depend on metal content are areas of ongoing research \citep{zhukovska08, galliano18}. It has been theorized that the smallest dust grains should be more susceptible to destruction by ionizing radiation in low metallicity environments due to their reduced heat capacities \citep{micel10a, micel10b, D21}. A deficiency of PAH emission has been observed in extreme environments such as \textsc{Hii} regions \citep{cesarsky96, kassis06, povich07, gordon08, chaste19, egorov23}, likely due to strong radiative processing. Therefore we expect in some cases it is possible for PAHs to be destroyed while larger dust grains survive. It is likely that grain formation and destruction processes shape the observed spectrum, and these may be more or less dominant in different environments and galaxies.

\vspace{1in}
\section{Data}\label{sec:data}

The primary data products used in this study are co-spatial Spitzer-IRS Short-Low (SL1 7.5~--~14.7\,\micron, SL2 5.2~--~7.6\,\micron, and SL3 7.4~--~8.7\,\micron) and Long-Low (LL1 20.6~--~38.4\,\micron\ and LL2 14.2~--~21.1\,\micron) spectroscopic maps. We extracted spectra from these maps in $\sim$0.43~--~1.7\,kpc apertures across three nearby galaxies from the PAH-BIGMAP program 20518: M101 (NGC~5457), NGC~628, and NGC~2403. These maps extend radially outwards from the center of each galaxy at an orientation designed to avoid bright \textsc{Hii} regions.  We also include eight Spitzer spectral observations of \textsc{Hii} regions in M101 from \citet{gordon08}, re-reduced in a consistent manner.

Figure~\ref{figure:prettyrgb} shows the position of each spectroscopic map on a three-color image of each galaxy. These images are composed of IRAC 3.6\,\micron\ in blue, IRAC 8\,\micron\ in green, and MIPS 24\,\micron\ in red (see \S\,\ref{sec:photdata}). Note the shift from green to red with increasing radius in each galaxy. This suggests the PAH emission at 8\,\micron\ is decreasing faster than the dust continuum emission at 24\,\micron. 

\begin{figure*}
\centering
\includegraphics[width=\linewidth]{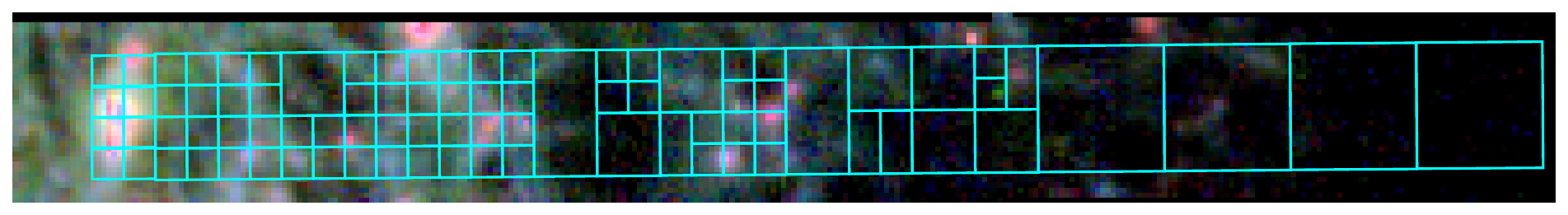}
\includegraphics[width=0.49\linewidth]{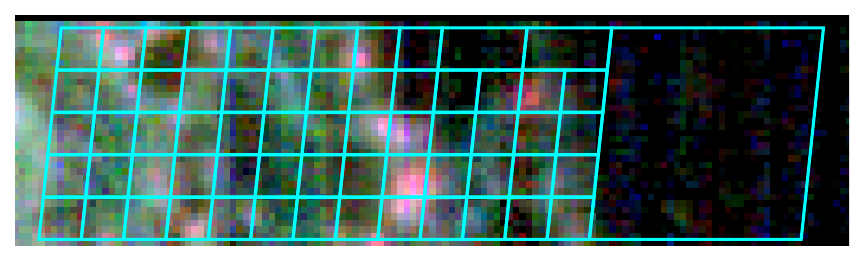}
\includegraphics[width=0.5\linewidth]{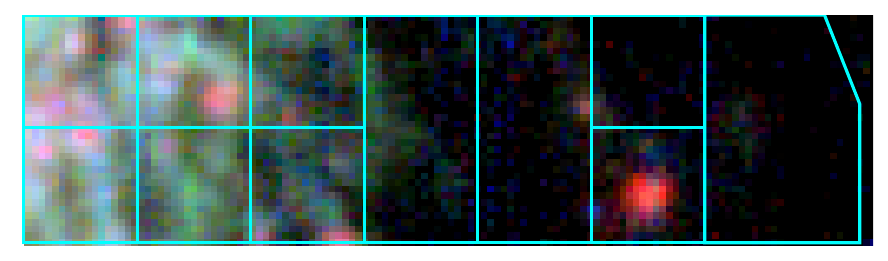}
\caption{\textbf{Three-color images from slices of the Spitzer-IRS SL1 spectral cubes for M101 (top), NGC~628 (left), and NGC~2403 (right).} Red is the surface brightness at the PAH 12.6\,\micron\ peak, green is at the PAH 11.3\,\micron\ peak, and blue is at the PAH 7.7\,\micron\ peak. The apertures defined in \S\,\ref{sec:regions} are shown in cyan.}
\label{figure:myrgbstrips}
\end{figure*}

\subsection{Spectroscopic Data Reduction}
All spectral maps were reduced with CUBISM \citep{jd07}, using background frames created from both the dedicated background spectra associated with each observation set, as well as emission-free regions on the outskirts of the radial strip maps.   Automatic and by-eye pixel selection was employed to remove bad pixels and mitigate striping in the final cube.

Slight mismatches in the surface brightness of aperture-matched spectra were found in faint regions for sub-orders SL1, SL2, and SL3. These were determined to arise at low surface brightness levels from light contamination which varies spatially on the detector and over time. We corrected the effect with a row-by-row box-car average of inter-order pixels across time-adjacent background-subtracted detector images. This process is implemented with the program \texttt{sl\textunderscore io\textunderscore correct} \citep{sandstrom12}. Correction using inter-order light results in a significant improvement in the flux agreement between SL1, SL2, and SL3 intensities at the wavelengths where they overlap for low surface brightness spectra.

The sub-orders of SL and LL were stitched by averaging the brightness at wavelengths where they overlap. The combined LL spectrum was then scaled to match the SL spectrum. This was done with a multiplicative factor or an additive offset based on the brightness of the spectrum. After removing backgrounds, residual baseline variations in dark regions of the cube remained below 1 MJy/sr.  For faint spectra (i.e. average brightness $<$0.75 MJy/sr), we employed an additive offset to bring spectral segments into agreement.  For brighter spectra, we employ a multiplicative factor.  We find this process results in good agreement between SL and LL for all spectra in our sample and all offsets and factors are small, typically $\sim$0.1 MJy/sr (additive) and $\sim$1.1 (multiplicative).

\subsection{Infrared Photometric Data}\label{sec:photdata}
We also compiled infrared photometric data from 8\,\micron\ through 160\,\micron\ in order to derive the total infrared (TIR) surface brightness in each of our apertures. The 8\,\micron\ data from the Infrared Array Camera (IRAC), the 24\,\micron\ data from the Multiband Imaging Photometer for Spitzer (MIPS), and the 70, 100, and 160\,\micron\ data from the Photodetector Array Camera and Spectrometer (PACS) of the Herschel Space Observatory were retrieved from the Key Insights on Nearby Galaxies: a Far-Infrared Survey with Herschel \citep[KINGFISH,][]{KINGFISH} and NGC~2403 from the Very Nearby Galaxies Survey \citep[VNGS,][]{VNGS}. NGC~2403 was never observed with PACS\,100\,\micron, so data at 250\,\micron\ from the Spectral and Photometric Imaging Receiver (SPIRE) of Herschel were included so the spectral energy distribution (SED) could be modeled.
All photometric data were convolved to the point-spread function (PSF) of the lowest resolution images used for SED determination -- PACS 160\,\micron, full-width half-maximum (FWHM) 11\farcs2, and a Gaussian PSF with FWHM 25\arcsec\ for NGC~2403.

\subsection{Region Definition}\label{sec:regions}
The spectral cubes for M101 and NGC~628 were convolved to the resolution of PACS 160\,\micron\ (11\farcs2) with kernels produced by the STINYTIM program \citep{sandstrom09}. Since NGC~2403 has no PACS 100\,\micron\ data, it was necessary to convolve the cubes to a Gaussian PSF of FWHM 25\arcsec\ in order to include longer wavelength far-infrared photometry in the SED fit.  The convolution process (when the image is padded with zeros) also dims the edges of each spectral cube where the convolution kernel falls off the coverage footprint. We trimmed the edges of our spectral cube by 5\arcsec\ on all sides to preserve as much area as possible while ensuring appropriate data coverage. This 5\arcsec\ is approximately the size of one pixel in the LL cubes, and just less than two pixels in the SL cubes.

Extraction apertures were defined to span the entire overlap region between the SL and LL footprints. We tiled each galaxy with rectangular apertures with a minimum side length of the limiting PSF of the supplemental photometric data (11\farcs2, 25\arcsec\ for NGC~2403). An iterative algorithm was used to define apertures based on the continuum brightness at MIPS 24\,\micron\ to ensure sufficiently high signal-to-noise PAH detections. The final apertures are shown in Figure~\ref{figure:myrgbstrips}. The initial apertures for this process were square and spanned the width of each radial strip. If the average surface brightness of the aperture was above 0.25 MJy/sr at MIPS 24\,\micron, then the square was split into a vertical pair and a horizontal pair. Next, we checked whether the vertical or horizontal pair was brighter. If the brightest pair was at least 10\% brighter than the original square, or above 1 MJy/sr, then the algorithm continued on the pair. The pair was then broken into two squares, and the same process was repeated until the minimum size is reached.

From all radial strips there are a total of 160 regions that meet the minimum continuum brightness threshold: 85 in M101, 64 in NGC~628, and 11 in NGC~2403. Including the 8 \textsc{Hii} regions, a total of 168 regions with well-measured spectra were analyzed.

\vspace{1in}
\subsection{MIR Spectral Fitting}\label{sec:pahfit}

The final extracted Spitzer spectra (5--38\,\micron) were fit with the IDL program PAHFIT \citep{jd07}. From PAHFIT we obtained the integrated intensities of all emission features and their propagated uncertainties. We do not expect significant attenuation in spectra of the spiral galaxies in our sample, so attenuation was not included in the fit. The sub-features of major PAH complexes were combined with PAHFIT to produce combined power in the 7.7, 11.3, 12.6, and 17\,\micron\ bands. We further combined all PAH features (5--18\,\micron) into \totpah\ using PAHFIT.

The H\textsc{ii} region spectral cubes are smaller and sparsely sampled by the slit, so we did not convolve them to a common resolution. Instead, we scaled our derived PAH luminosities in these apertures by the ratio between their average native IRAC 8\,\micron\ photometry and their average IRAC 8\,\micron\ when convolved to PACS 160\,\micron\ resolution. The median ratio is 1.94, the mean is 1.86, and the standard deviation is 0.25, so the PAH luminosity of each H\textsc{ii} region was scaled down by about a factor of two, with no change in the band ratios.

\begin{figure*}
\includegraphics[width=\linewidth]{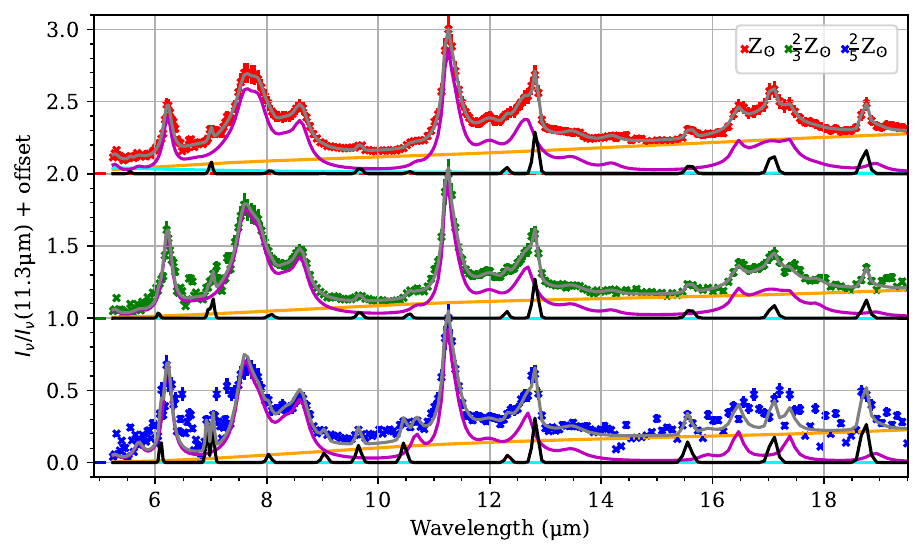}
\caption{\textbf{Example IRS spectra from M101.} The spectrum of the center (red), threshold metallicity (green), and edge regions (blue) from the M101 radial strip. Each spectrum is normalized by the surface brightness at 11.3\,\micron. The PAHFIT result for each spectrum is shown in gray along with each component of the fit: stellar continuum (cyan), dust continuum (orange), line emission (black), and PAH emission (magenta). Note the relative shift in power from long wavelength PAH features to short as metallicity drops from \zsolar\ to 0.4\,\zsolar.}
\label{figure:examplespectra}
\end{figure*}

\begin{figure*}
\centering
\includegraphics[width=0.9\linewidth]{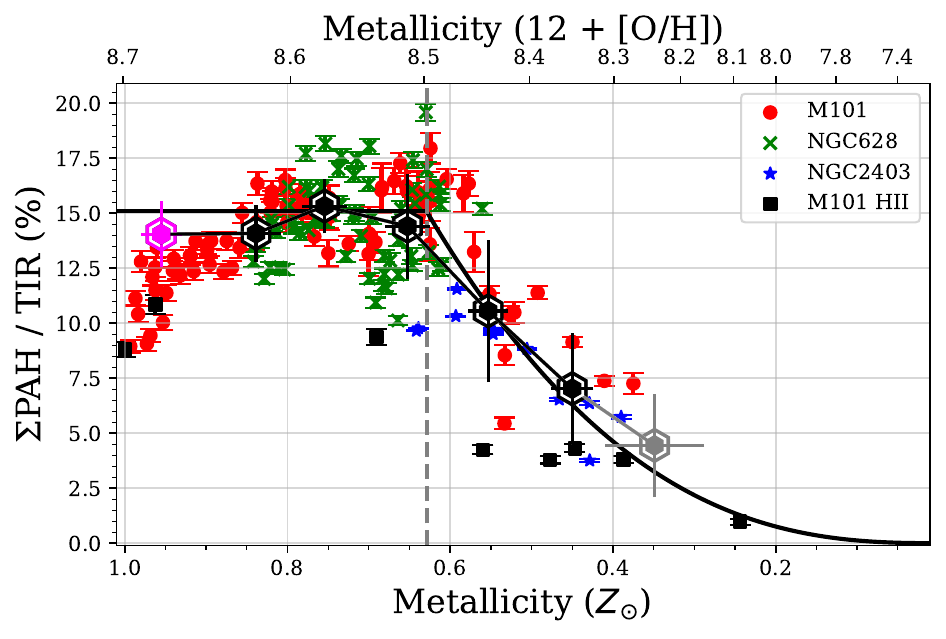}
\includegraphics[width=0.9\linewidth]{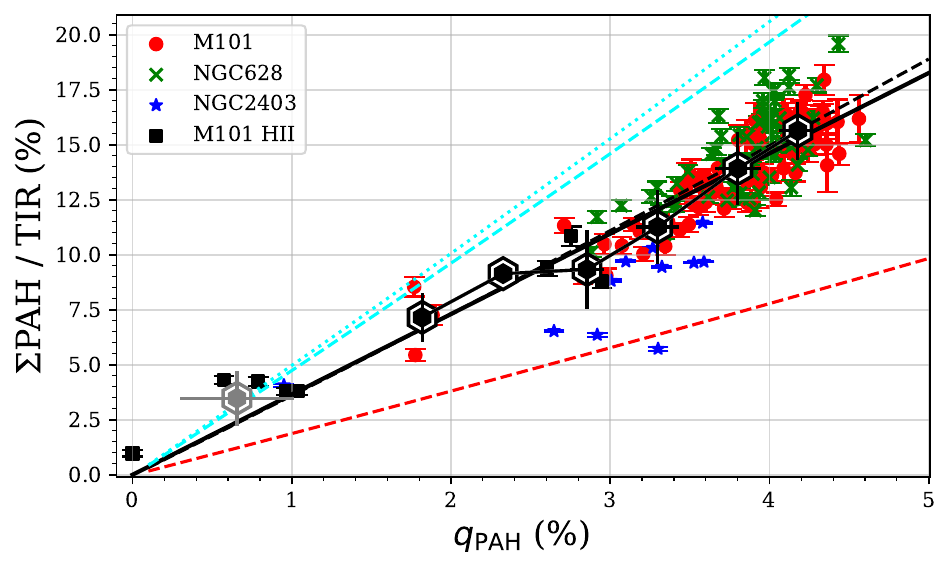}
\caption{\textbf{(Top) PAH emission deficit at low metallicity.} Ratio of all PAH emission to all other dust emission (\pahtir) as a function of metallicity. The hexagons and their errorbars indicate the mean and standard deviation in six bins of width 0.1\,\zsolar. The gray bin includes only the points below 0.4\,\zsolar. The pink bin falls within the bulge radius of M101 and indicates the mean after a correction for reddened ISRF is applied (see \S\,\ref{sec:bulge}). The solid black lines indicate the median above \zthresh\ and a power law fit below. \textbf{(Bottom) Correlation between \pahtir\ and modeled PAH mass abundance fraction (\qpah).} The dashed lines indicate the expected trend from D21 models for three incident ISRFs with $U$=1: M31 bulge (red), mMMP (black), and a low-metallicity starburst (cyan). The dotted cyan line indicates the low-metallicity spectrum with $U$=100. The solid black line indicates a linear fit with intercept fixed at zero. The hexagons indicate the mean and standard deviation in bins of width 0.5\% in \qpah. The gray bin includes only the points below \qpah~=~1\%.}
\label{figure:pahtir}
\end{figure*}

\subsection{Derived Quantities}\label{sec:eqns}
From the photometric data we derived total infrared luminosity (3--1100\micron, TIR) for each region in M101 and NGC~628 using the calibration presented in \citet{galam13}:
\begin{equation}
\begin{split}
    \rm TIR &~=~ 2.051 \times \nu I_{\nu} (24\micron)\,+ 0.521 \times \nu I_{\nu} (70\micron)~\\
    &~~\,+ 0.294 \times \nu I_{\nu} (100\micron) + 0.934 \times \nu I_{\nu} (160\micron)
\end{split}
\label{eqn:tir1}
\end{equation}
where the constants were taken from their Table 3 and $I_{\nu}$ ($\lambda$) is the surface brightness at $\lambda$ microns in MJy/sr.

There is no data at 100\,\micron\ for NGC~2403, so we used the calibration from \citet{galam13} that does not depend on this band:
\begin{equation}
\begin{split}
    \rm TIR &~=~ 2.126 \times \nu I_{\nu} (24\micron)~ + 0.670 \times \nu I_{\nu} (70\micron)\\
    &~~\,+ 1.134 \times \nu I_{\nu} (160\micron)
\end{split}
\end{equation}
where the notation is the same as Equation~\ref{eqn:tir1}.

We determined metallicities (\twelog) for our extracted regions in M101, NGC~628, and NGC~2403 using the radial gradient equations presented in \citet{chaosVI}. We defined solar metallicity as \zsolar~$\equiv$~\twelog~=~8.69 \citep{asplund09} and we assumed that O/H is a good tracer of the abundance of all metal species. \citet{chaosV} found the relative C/O abundance decreases radially in M101 by about 0.4 dex over two effective radii (R$_e$, see Table~\ref{table:gradtable}). Therefore there is likely even less carbon than expected from a linear scaling with the oxygen abundance gradient of each galaxy. The parameters of the metallicity gradient equation for each galaxy are shown in Table~\ref{table:gradtable}, where the metallicity is largest at the center (\zcent) and decreases with a constant slope ($m$) as a function of deprojected galactocentric radius ($R$): \zmetal~$=$~\zcent~$-~m~\times~(R/R_{\rm e})$.

We assumed metallicity is purely radial with no small scale or azimuthal variations. \citet{kreckel19} found the metallicity of nearby spiral galaxies varies by up to 0.05~dex on scales of 120~pc, but the radial dependence dominates. This is comparable to the scatter in the metallicity gradient fits employed here. Including a 0.05~dex uncertainty for all metallicity values would not significantly alter the results of this work.

\subsubsection{Dust and Galaxy Property Maps}\label{sec:SEDmodel}

The properties of dust and local environmental conditions can be derived from infrared photometric data from IRAC~3.6\,\micron, IRAC~4.5\,\micron, IRAC~5.8\,\micron, IRAC~8\,\micron, MIPS~24\,\micron, PACS~70\,\micron, PACS~100\,\micron, and PACS~160\,\micron. This was done by fitting the SED with the \citet{draineli07} model using the methodology of \citet{aniano20}. NGC~2403 was not observed with PACS 100\,\micron\ so an additional far-infrared band, SPIRE 250\,\micron\ was necessary to accurately fit the SED model. 
The SED models produce resolved maps of quantities such as the fraction of dust mass in the form of PAHs with \carbons~$<~10^{3}$ (\qpah), and the mass-weighted average starlight energy density per unit frequency that is illuminating the local dust ($\bar{U}$). We focus primarily on the maps of these two quantities in this analysis. Values for each region are extracted from these maps using the matched apertures defined in \S\,\ref{sec:regions}.

\begin{deluxetable}{cccc}
\tabletypesize{\scriptsize}
\tablecaption{Galaxy Information \& Metallicity Gradients \label{table:gradtable}} 
\tablewidth{0pt}
\tablehead{
\colhead{\textbf{Property}} &
\colhead{\textbf{M101}} &
\colhead{\textbf{NGC 628}} &
\colhead{\textbf{NGC 2403}} 
}
\startdata
Central Abundance (12~+~[O/H]) & 8.70~$\pm$~0.04 & 8.65~$\pm$~0.05 & 8.55~$\pm$~0.04 \\
Slope $m$ (dex/R$_e$) & 0.17~$\pm$~0.02 & 0.09~$\pm$~0.03 & 0.09~$\pm$~0.03 \\
R$_e$ (arcsec) & 197.6 & 95.4 & 178.0 \\
Distance (Mpc) & 7.4 & 7.2 & 3.2 \\
Inclination ($\degree$) & 18 & 5 & 63 \\
Position Angle ($\degree$) & 39 & 12 & 124 \\
\enddata
\tablenotetext{}{Values adopted from \cite{chaosIV} and \cite{chaosVI}.}
\end{deluxetable}

\section{Observational Results}\label{sec:obsresults}

Figure~\ref{figure:examplespectra} shows three example spectra from the M101 radial strip. Comparing them, we observe the 17\,\micron\ feature and the brightest feature at 7.7\,\micron\ are significantly reduced at lower metallicity. The remaining PAH power is in the 11\,\micron\ complexes and the 6.2\,\micron\ feature which is brighter at 0.4\,\zsolar\ than at \zsolar.

\subsection{PAH to Total Dust Emission Metallicity Relation}

The ratio of all PAH emission (\totpah) to the total infrared brightness (TIR) is shown in the top panel of Figure~\ref{figure:pahtir}. \pahtir\ rises from about 10\% to 16\% as metallicity decreases from \zsolar\ to $\sim\frac{2}{3}$\,\zsolar, then drops below 2\% by $\sim\frac{1}{4}$\,\zsolar. We refer to this threshold metallicity where the drop in \pahtir\ occurs as \zthresh~$\equiv$~0.625\,\zsolar\ (\twelog~=~8.5). The best-fit relation between \pahtir\ and metallicity has the form:

\begin{equation}
\rm \frac{\Sigma PAH}{TIR} (\%)  =
\begin{cases}
    15.1~\pm~2.0 &~~Z > Z_\mathrm{th}\\
    (15.1 \pm~0.2) \left(\frac{Z}{~Z_{\rm th}}\right)^{2.6~\pm~ 0.1}&~~Z \leq Z_\mathrm{th}
\end{cases}
\end{equation}

\noindent where the fit at metallicities above \zthresh\ is the uncertainty-weighted median with bulge light correction applied (see \S\,\ref{sec:bulge}) and below \zthresh\ the fit is a power law. The value of \zthresh\ is determined by requiring the fit to be continuous at \zthresh\ to within 1$\sigma$. The \textsc{Hii} regions in the mid-disk are notably offset from the overall trend; their \pahtir\ is somewhat lower than expected based on their metallicity.

The bottom panel of Figure~\ref{figure:pahtir} shows the strong correlation between \pahtir\ and \qpah, where the linear fit has the form:

\begin{equation}
\begin{split}
    &\rm \frac{\Sigma PAH}{TIR} = (3.69~\pm~0.02) \times \textit{q}_{PAH}
\end{split}
\end{equation}

Also indicated in this figure is the expected \pahtir\ for a given \qpah\ from the D21 models. We use their `standard' GSD and re-scale it such that the resulting integral for \carbons$~<~10^3$ gives a desired \qpah. A spectrum is then produced from each GSD and \pahtir\ is calculated by PAHFIT and Equation~\ref{eqn:tir1}. We leave $U$ fixed at 1 and repeat this using three different ISRFs to plot the three dashed lines: the 10 Myr \zsolar/20 SED from \citet{bpass}, the solar neighborhood mMMP SED from \citet{mmp83, D21}, and the M31 bulge SED from \citet{groves12}. An additional dotted line is also shown where we have increased $U$ to 100 and used the 10 Myr \zsolar/20 SED. These lines indicate the general effect that increasing the ISRF hardness and intensity has on the PAH and dust luminosities.

\subsubsection{Bulge ISRF Correction}\label{sec:bulge}

\begin{figure}
\includegraphics[width=\linewidth]{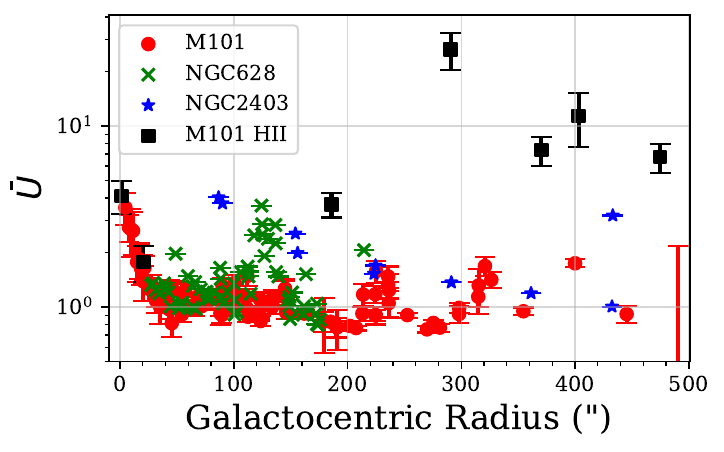}
\caption{\textbf{$\bar{U}$ as a function of galactocentric radius.} Values detected below 3$\sigma$ are set to zero.}
\label{figure:Ubar}
\end{figure}

Following the work of \citet{andromeda}, we can interpret the decrease in \pahtir\ as metallicity increases from 0.9\,\zsolar\ to \zsolar\ as a result of the increasing bulge fraction at low galactocentric radii in M101. NGC~628 and NGC~2403 do not have significant bulges \citep{fisher10}. PAHs in the bulge are excited by an ISRF dominated by older stars, so they emit less compared to PAHs in the disk where the ISRF is more dominated by UV and optical photons from young stars. This effect is less significant for larger dust grains, so \pahtir\ drops as the bulge fraction increases. The red dashed line in Figure~\ref{figure:pahtir} shows this using the D21 models: a given \qpah\ illuminated by the M31 bulge SED results in $\sim50\%$ less \pahtir\ than the same \qpah\ illuminated by the solar neighborhood mMMP SED.

Figure~\ref{figure:Ubar} shows the $\bar{U}$ values extracted from our apertures. Based on this, the disk of M101 has $\bar{U}$~$\approx$~1 and it begins to increase due to the bulge component of the galaxy closer to the center. We consider the bulge radius to be where $\bar{U}$ begins to increase above 1 (about 60$\arcsec$ from galaxy center, approximately where metallicity is 0.9\,\zsolar). Within this radius is also where the drop in \pahtir\ and \qpah\ occurs for regions in M101.

We used the equations from \citet{andromeda} to correct \qpah\ by accounting for the softer radiation field in the bulge of M101. Again using the $\bar{U}$ maps we calculated $U_{bulge}$  for all regions within the bulge radius as $U_{bulge}$ = $\bar{U}$ - $U_{disk}$ with $U_{disk}$ = 1. However we find that this shifts points within the bulge radius from 1\% below the median \qpah\ between \zthresh\ and 0.9\,\zsolar\ (4\%) to 1\% above the median. A correction that brings these points to share the median can be obtained by a small modification to Equation~21 from \citet{andromeda}. The parameter $A$ is tied to the hardness of the radiation field with higher values corresponding to harder radiation. For instance the $A$ value for the bulge of M31 is 1.95 and the value for the disk is 4.72. We find changing only the value of $A_{bulge}$ from 1.95 to 3.85 brings the points in the bulge radius to agree with the median of the points outside this radius (and less than \zthresh) in both \pahtir\ and \qpah. This brings $A_{bulge}$ closer to $A_{disk}$~=~4.72. This change is reasonable since M101 has a weak, actively star-forming bulge while the bulge of M31 considered in \citet{andromeda} has over three times the metal abundance, significantly more dust attenuation, and an older stellar population dominating the ISRF.

\subsection{PAH Band Ratios}\label{pahratios}

\begin{figure}
\includegraphics[width=\linewidth]{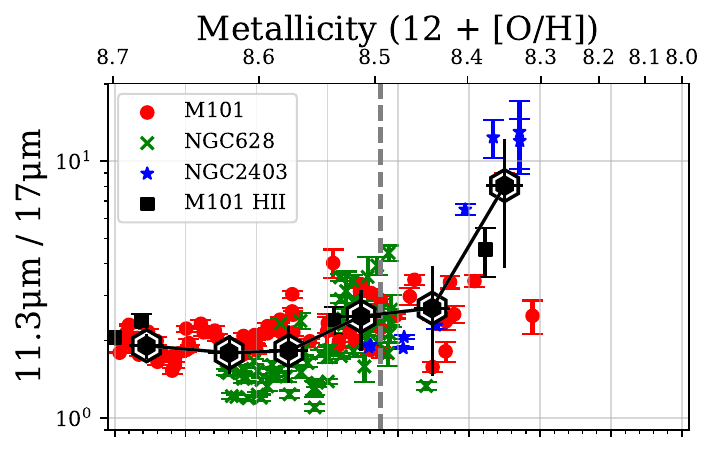}
\includegraphics[width=0.985\linewidth]{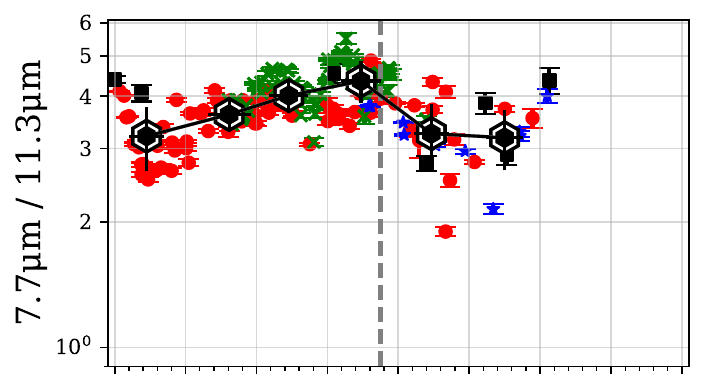}
\includegraphics[width=0.985\linewidth]{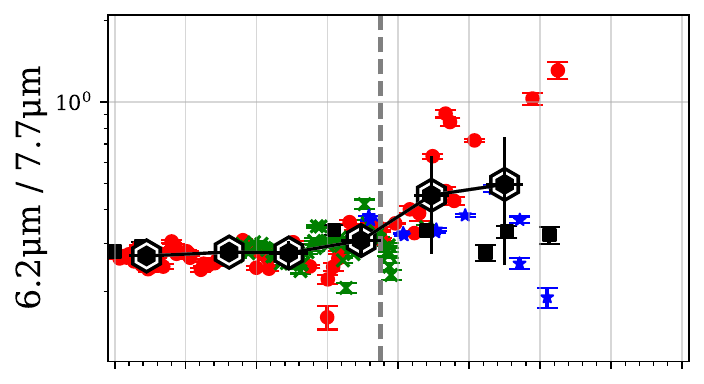}
\includegraphics[width=\linewidth]{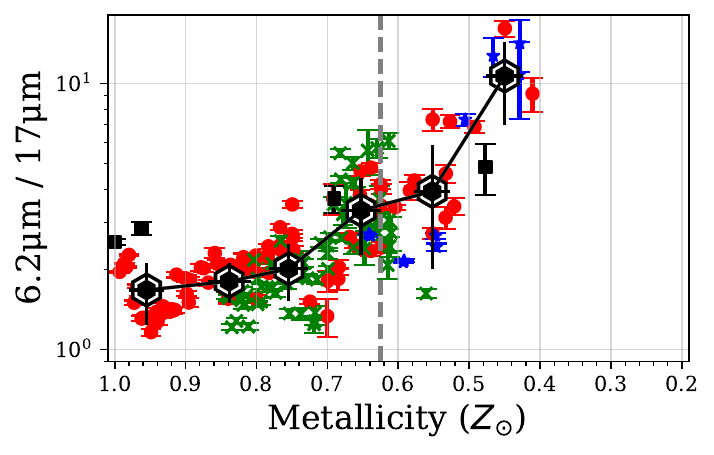}
\caption{\textbf{As a function of metallicity: (top) \elevteen\ ratio, (second) \sevelev\ ratio, (third) \sixsev\ ratio, (bottom) \sixteen\ ratio.} Hexagons indicate the mean and standard deviation in six bins with width 0.1\,\zsolar. The vertical line indicates the threshold metallicity \zthresh. Ratios of features that were detected below 3$\sigma$ are not included.}
\label{figure:ratios}
\end{figure}

To better diagnose what is happening with the PAHs, we plot short-to-long wavelength band ratios as a function of metallicity in Figure~\ref{figure:ratios}. The top panel shows the ratio \elevteen\ is constant until \zthresh, then it rises steeply as metallicity drops. Since the 17\micron\ feature originates from the largest neutral and ionized PAHs, this could be evidence that the largest PAHs ($>~2,000~$\carbons) are preferentially depleted at low metallicity. This rise could also be a result of increasing ISRF hardness as metallicity decreases below \zthresh\ causing a shift in PAH luminosity from longer to shorter wavelength bands \citep{D21}.

The second panel shows the ratio \sevelev\ decreases away from the threshold metallicity on either side. The 7.7\,\micron\ feature is emitted most by ionized PAHs with $\sim100$~\carbons\ and the 11.3\,\micron\ feature is from neutral PAHs with $\sim400$~\carbons. As metallicity decreases from \zsolar\ to \zthresh\ the PAHs may be getting smaller or more ionized, then below \zthresh\ they may become larger or less ionized. It is also possible that the harder ISRF at lower metallicity is raising the fractional PAH luminosity at short wavelengths and resulting in an increase in the \sevelev\ ratio, however this could not explain why the ratio then decreases below \zthresh.

The third panel shows the ratio \sixsev\ is mostly constant from \zsolar\ to \zthresh, then increases at lower metallicities. The increase is steeper for the M101 strip regions than for those from NGC~2403 and the M101 \textsc{Hii} regions. The 6.2\,\micron\ feature is from ionized PAHs with $\sim70$~\carbons\ so the increase at low metallicity could be a result of PAHs becoming smaller below \zthresh. It is also possible that the increasing ISRF hardness is responsible for the increase in the \sixsev\ ratio.

The final panel shows the ratio \sixteen\ rises steadily as metallicity decreases. This again indicates that the PAHs are becoming smaller on average. Alternatively the PAHs may remain the same and the ISRF hardness is increasing, causing the PAHs to emit more at shorter wavelengths.

Overall the trends in these four PAH band ratios suggest the following. \sevelev\ and \sixteen\ rise between \zsolar\ and \zthresh\ while \sixsev\ and \elevteen\ are constant. This suggests the effect of increasing ISRF hardness is not significant at these metallicities, or else all ratios would show a steady increase. The observed increase in \sixteen\ by about an order of magnitude could be interpreted as an increase in the amount of PAHs with $\sim70$~\carbons\ relative to PAHs with $\sim2,000$~\carbons. Similarly, the observed increase in \sevelev\ by about 40\% could be interpreted as an increase in the amount of PAHs with $\sim100$~\carbons\ relative to PAHs with $\sim400$~\carbons.

It is also possible that the increase in \sevelev\ is evidence of PAHs becoming more ionized. This is the only ratio shown that is strongly dependent on the ionization distribution of the PAHs. The ionization fraction of PAHs depends on the conditions of the gas and the grain size, with smaller PAHs being more neutral on average than larger PAHs \citep{D21}. If PAHs are becoming smaller then the overall fraction of neutral PAHs will increase as well.

Below \zthresh, again all ratios except for \sevelev\ show the same trend of increasing with decreasing metallicity. The three ratios that increase at low metallicity suggest PAHs overall are becoming smaller and/or hotter. If changes in the radiation field hardness were the primary driver, we would expect the \sevelev\ ratio to also increase below \zthresh.  This is not what is observed, suggesting instead that there are also changes in the PAH size distribution and overall ionization fraction.

\begin{figure}
\includegraphics[width=\linewidth]{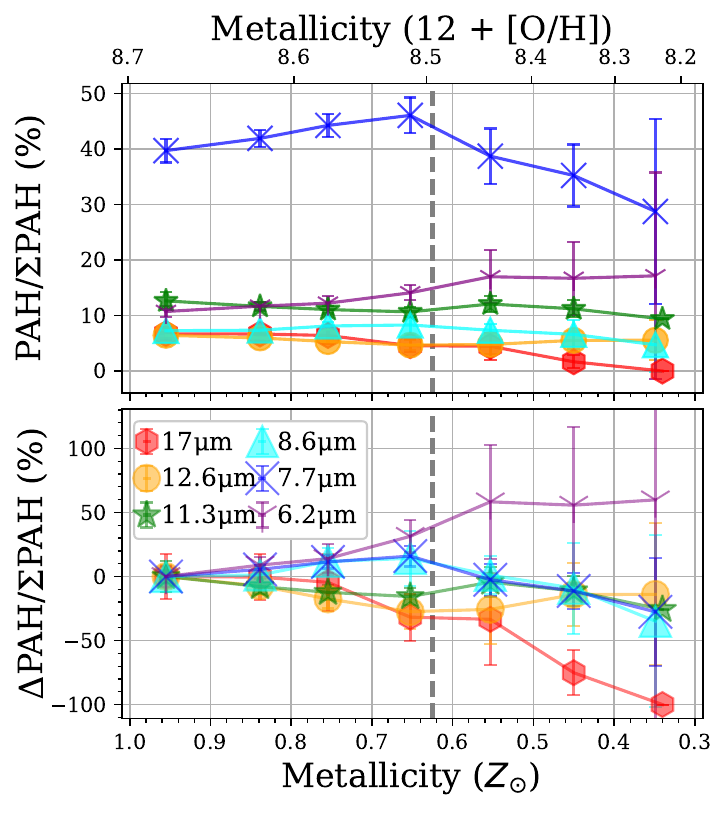}
\caption{
\textbf{(Top) The observed metallicity effect on PAHs from the ratio of each major PAH feature to total PAH luminosity.} The gray vertical line indicates the threshold metallicity \zthresh. \textbf{(Bottom) Fractional change in \rattot\ relative to the value at \zsolar.} Relative to the total PAH luminosity, the power shifts to shorter wavelengths, then after \zthresh\ the 7.7 and 17\,\micron\ features decrease, the 11.3\,\micron\ feature is unchanged, and the 6.2\,\micron\ feature increases.
}
\label{figure:levers}
\end{figure}

The bottom panel of Figure~\ref{figure:levers} shows the differential change in each PAH feature relative to the total PAH luminosity compared to the value at solar metallicity. This figure shows the shifts in power among the PAH features. As metallicity drops from \zsolar\ to \zthresh, the PAH power shifts out of the longer wavelength features and into the short wavelength features. When metallicity drops below about \zthresh\ the power begins to shift out of the 7.7\,\micron\ PAH feature and toward the 11-13\,\micron\ complexes. The most dramatic changes are in the 6.2\,\micron\ and 17\,\micron\ feature. The 6.2\,\micron\ feature increases about 60\% from \zsolar\ to $\frac{1}{2}$\,\zsolar. The 17\,\micron\ feature decreases about 75\% over this same metallicity range.

\section{Modeled PAH \& Dust SED Analysis}\label{sec:dvresults}

\begin{figure}
\centering
\includegraphics[width=\linewidth]{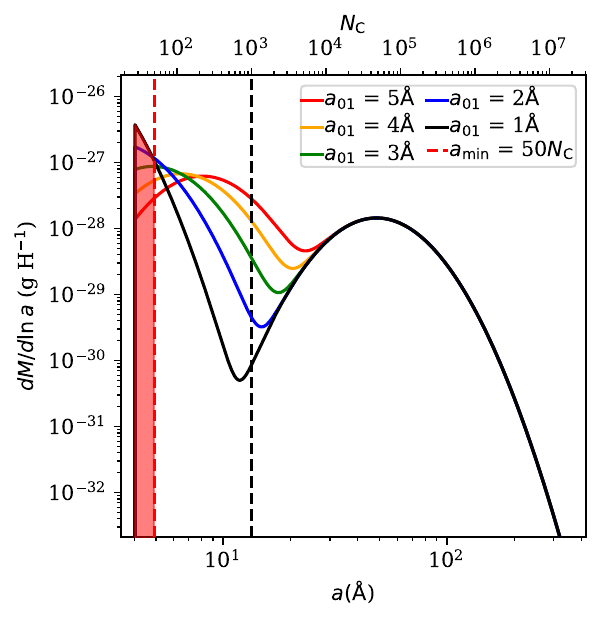}
\caption{\textbf{Size distributions of PAH feature emitting grains.} Shown are the size distributions as described in \citet{D21} with several values of \aone. Note that changing \aone\ shifts the location of the first peak but the integral remains constant. The photo-destruction parameter $a_\mathrm{min}$ is shown in red; below this value all PAHs are deleted from the distribution without preserving the integral.}
\label{figure:sizedist}
\end{figure}

To interpret the changes in band ratios, we created models of PAHs with various size distributions and illuminated by different radiation fields using the theoretical PAH spectra from D21. By comparing the variations in \qpah, \pahtir, and \rattot\ we investigated what changes in PAH size and ISRF can reproduce our observed trends. D21 produced model PAH spectra based primarily on four input parameters: incident radiation field spectrum, incident radiation intensity ($U$), PAH grain size distribution, and a binary PAH ionization distribution (either neutral or cationized). The standard grain size distribution is described in D21 as a double-peaked log-normal distribution with peak locations characterized by grain sizes \aone$=4$\,\AA\ and $a_{02}=30$\,\AA\ (see Figure~\ref{figure:sizedist}). 
Based on the agreement between the linear fit in the bottom panel of Figure~\ref{figure:pahtir} and the D21 model using the mMMP ISRF with $U~=~1$, we assumed these parameters are constant in each of the scenarios described below other than the ISRF hardness scenario. For simplicity, we used the standard ionization distribution as a function of grain size. When we generated an infrared dust SED from the D21 models, we also included the emission from larger `astrodust' grains in order to get an accurate measure of TIR.

For each spectrum we generated from the D21 models, we measured all the PAH features using PAHFIT \citep{jd07} to be consistent with the PAH feature strengths measured in the observed spectra. We used an updated version of the PAHFIT model that includes the 3.3\,\micron\ PAH feature \citep{lai20}. From the PAHFIT results we derived the sum of all PAH emission (\totpah) as well as the ratio of every PAH band relative to \totpah\ (\rattot). TIR was determined by integrating the generated spectrum from 3 to 1,100\,\micron. \qpah\ for every spectrum is calculated by the ratio of PAH mass -- from integration of the input GSD for grains with \carbons~$<~10^3$ ($\sim13.4\,\mathrm{\AA}$) -- relative to the total astrodust mass.

\subsection{Model Scenarios}

We considered three realistic scenarios that could reproduce the major observed PZR trends: \qpah, \pahtir, and the main \rattot\ ratios, all as a function of metallicity. 
The `ISRF hardness' model fixes the size distribution and varies the hardness of the incident ISRF as metallicity decreases. The `photo-destruction' model simulates bottom-up destruction of PAHs by increasing the minimum grain size as metallicity decreases. Finally, the `inhibited growth' model mimics a reduced accretion rate of carbon from the gas phase by decreasing the average PAH size and the overall abundance of carbonaceous grains.

\subsubsection{ISRF Hardness Model}\label{sec:hardmodel}

\begin{figure}
\centering
\includegraphics[width=\linewidth]{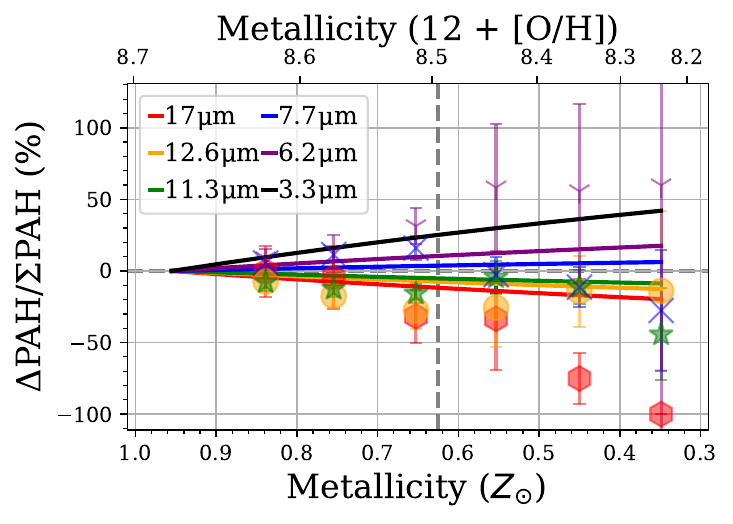}
\caption{
\textbf{Fractional PAH power as a function of metallicity for the ISRF hardness model (\S\ref{sec:hardmodel})}. The binned data from Figure~\ref{figure:levers} are shown. Also shown is the model-predicted trend for the fractional change in the relative 3.3\,\micron\ feature brightness (black line).
}
\label{figure:hardlever}
\end{figure}

First, we test if changes in the incident ISRF SED alone can explain the observed PZR trends. For this test we use the two ISRF hardness extremes: the bulge SED of M31 and a 10 Myr \zsolar/20 SED (low-Z). At high metallicity (at galaxy center), we assume the ISRF is weighted as 1:0 M31:low-Z, and at the lowest metallicity (at galaxy edge) we assume the weighting is 0:1 M31:low-Z with a linear interpolation between these extremes. The low-Z SED is characteristic of significantly lower metallicity than we expect at the edges of our three galaxies, and the M31 bulge SED is likewise significantly softer than we expect in the centers of our galaxies. Therefore we expect that the resulting changes to the PAH ratios from this ISRF hardness model represent an upper bound and more realistically the radial change in radiation hardness for our three galaxies will be smaller than suggested by this model.

Figure~\ref{figure:hardlever} shows the effect of changing the ISRF hardness on the ratios \rattot: trends broadly similar to the observations are produced, but with a magnitude that is insufficient by a factor of 3--5. Since the ratios from this model represent an upper bound, this indicates that the increased radiation hardness at low metallicity may play a role in the observed PZR trends, but the effect is not strong enough to be the primary driver. This model is also incapable of producing the observed drop in \qpah\ and \pahtir\ as metallicity decreases below \zthresh. When only the incident radiation field hardness is increased, PAHs of all sizes emit more, so \totpah\ increases. Due to the difference in heat capacities between PAHs and larger dust grains that emit mid- and far-IR continuum, \totpah\ increases faster than TIR in harder radiation environments. Since \qpah\ is derived from the GSD and we do not change it in this scenario, \qpah\ remains constant regardless of the incident ISRF.  In practice \qpah\ is often inferred from the 8\,\micron\ brightness and, similar to \pahtir, this increases faster than the total dust luminosity as the incident radiation field becomes harder. Therefore changing the ISRF hardness alone cannot explain all observed PZR trends.

\subsubsection{Photo-Destruction Model}\label{sec:guillotine}

\begin{figure*}
\includegraphics[width=0.55\linewidth]{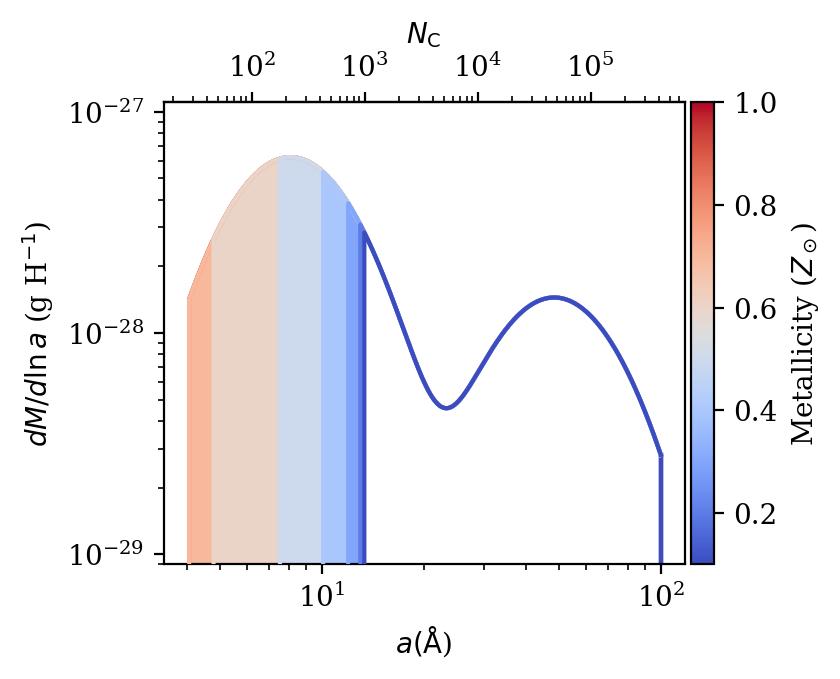}
\includegraphics[width=0.48\linewidth]{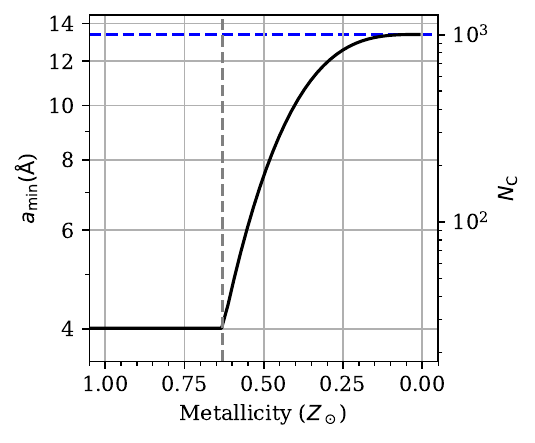}
\includegraphics[width=\linewidth]{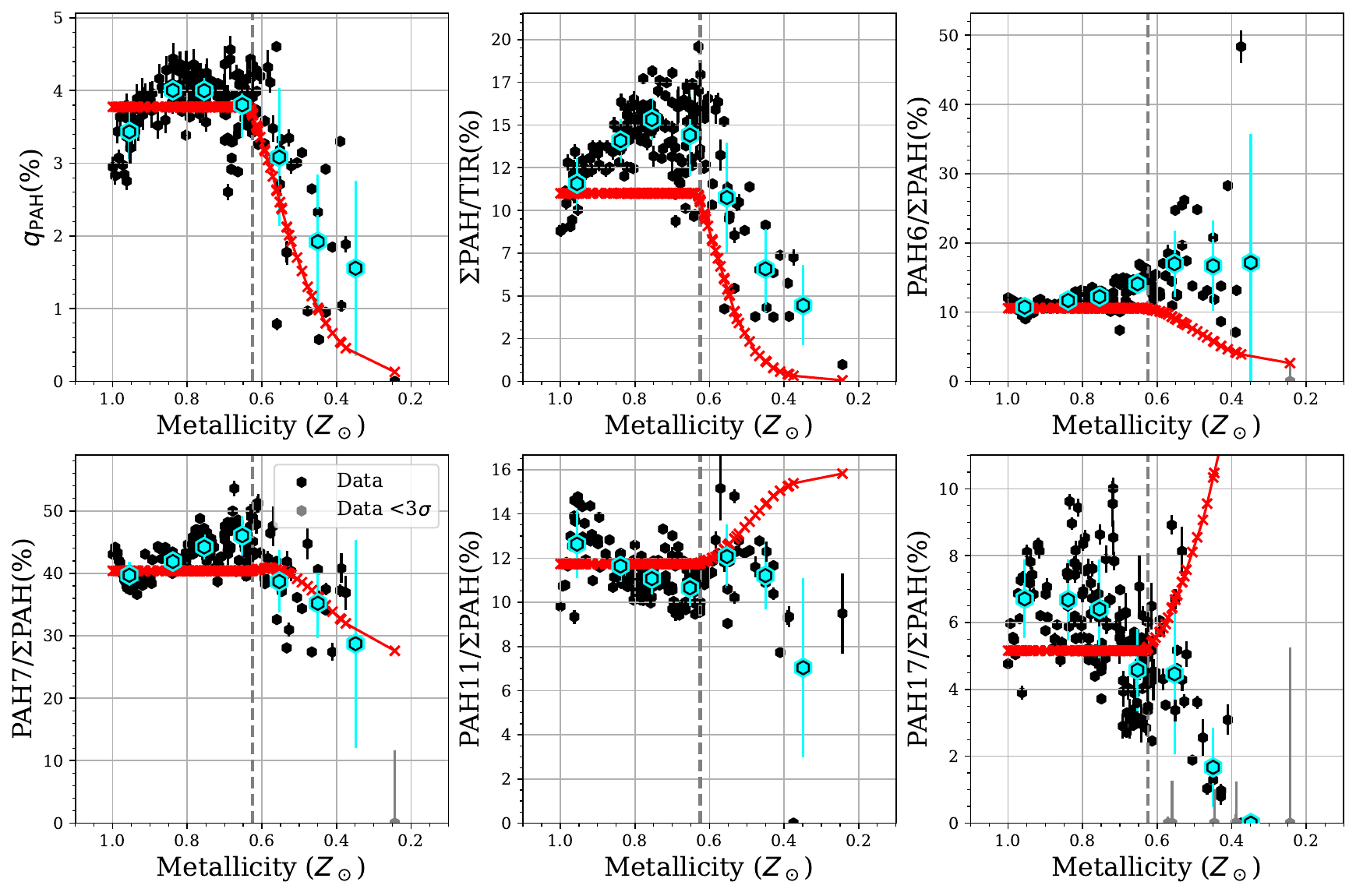}
\caption{\textbf{Grain photo-destruction model summary.} \textbf{Top-Left:} the size distribution of PAH-emitting grains varies as a function of metallicity, with the lower size cutoff $a_\mathrm{min}$ increasing as metallicity drops below a specific threshold, at a rate (\textbf{top-right}) which reproduces the \qpah-$Z$ relationship. The dashed blue line indicates grains with \carbons~=~10$^3$ so at this $a_\mathrm{min}$ \qpah\ = 0.
\textbf{Bottom}: comparisons between six observed PZR trends (black) and the photo-destruction model (red). The dashed vertical line indicates the threshold metallicity below which $a_\mathrm{min}$ begins increasing.  Seven metallicity bins indicate the median of each trend in cyan, with detections below 3$\sigma$ set to zero.}
\label{figure:guillosumm}
\end{figure*}


We simulate the effects of photo-destruction of small PAHs as the average hardness of the radiation field increases inversely with metallicity by imposing a sliding minimum grain size below which all PAHs are destroyed.  In this scenario, \qpah\ and \pahtir\ decrease because small PAHs are removed from the population starting from the smallest grains up. 
We model this with a minimum grain size cutoff, $a_\mathrm{min}$,  which rises as metallicity drops.  The functional form of $\log(a_\mathrm{min}(Z))$ is a power law, with index determined by minimizing $\chi^2$ between \qpah\ from the model GSDs and the observationally-determined \qpah\ values. The model is summarized in Figure~\ref{figure:guillosumm}. The best fit correlation for $a_\mathrm{min}$($Z$) is:

\begin{equation}
    \log\left(\frac{a_\mathrm{min}}{4 \rm \AA}\right) ~=~
\begin{cases}
    0 &~~Z > Z_\mathrm{th}\\
    0.52 \times \left[1 - \left(\frac{Z}{~Z_\mathrm{th}}\right)^{3.2}\right]&~~Z \leq Z_\mathrm{th}
\end{cases}
\label{eqn:guillotine}
\end{equation}

\noindent and is shown in the top right panel of Figure~\ref{figure:guillosumm}. We find this model can be fit to the \qpah\ trend and it coarsely reproduces the \pahtir\ decline. The resulting band ratio trends, however, are a poor match for the observations. Since \qpah\ is constant at metallicities higher than \zthresh, the modeled band ratios likewise do not change. We found this model does not explain many of the observed trends.  The resulting shift in the PAH band ratios with metallicity is for many features \emph{opposite} of the observed trend, as can be expected for a scenario which preferentially weakens short wavelength bands.
For instance, the bottom panel of Figure~\ref{figure:guillosumm} shows that modeled \sixtot\ decreases but the observed trend increases, and \teentot\ increases dramatically in the model, while the observed trend drops steeply.
The model does reproduce the drop in the 7.7\,\micron\ feature below \zthresh, however it does not explain the observed rise in this feature from \zsolar\ to \zthresh. Even a combination of the radiation hardness and photo-destruction models cannot produce the observed shift of power from long to short wavelength PAH features, because the small PAHs that emit most effectively at shorter wavelengths are being preferentially destroyed faster than the radiation hardness is increasing.

\begin{figure*}
\includegraphics[width=0.52\linewidth]{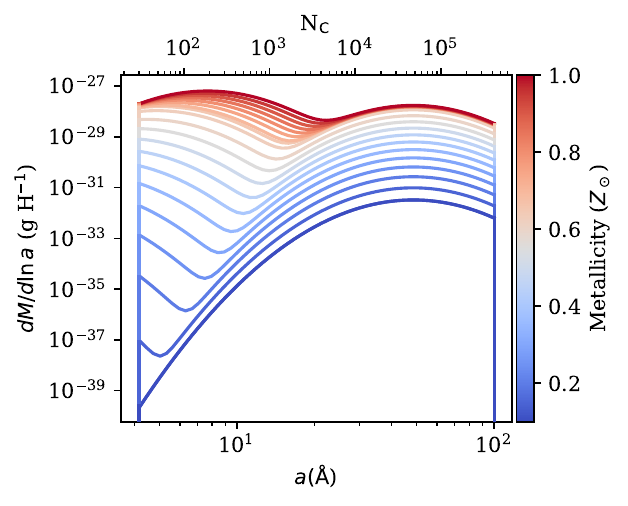}
\includegraphics[width=0.48\linewidth]{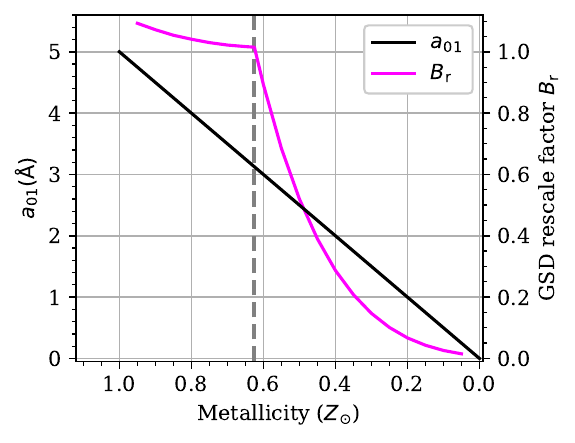}
\includegraphics[width=\linewidth]{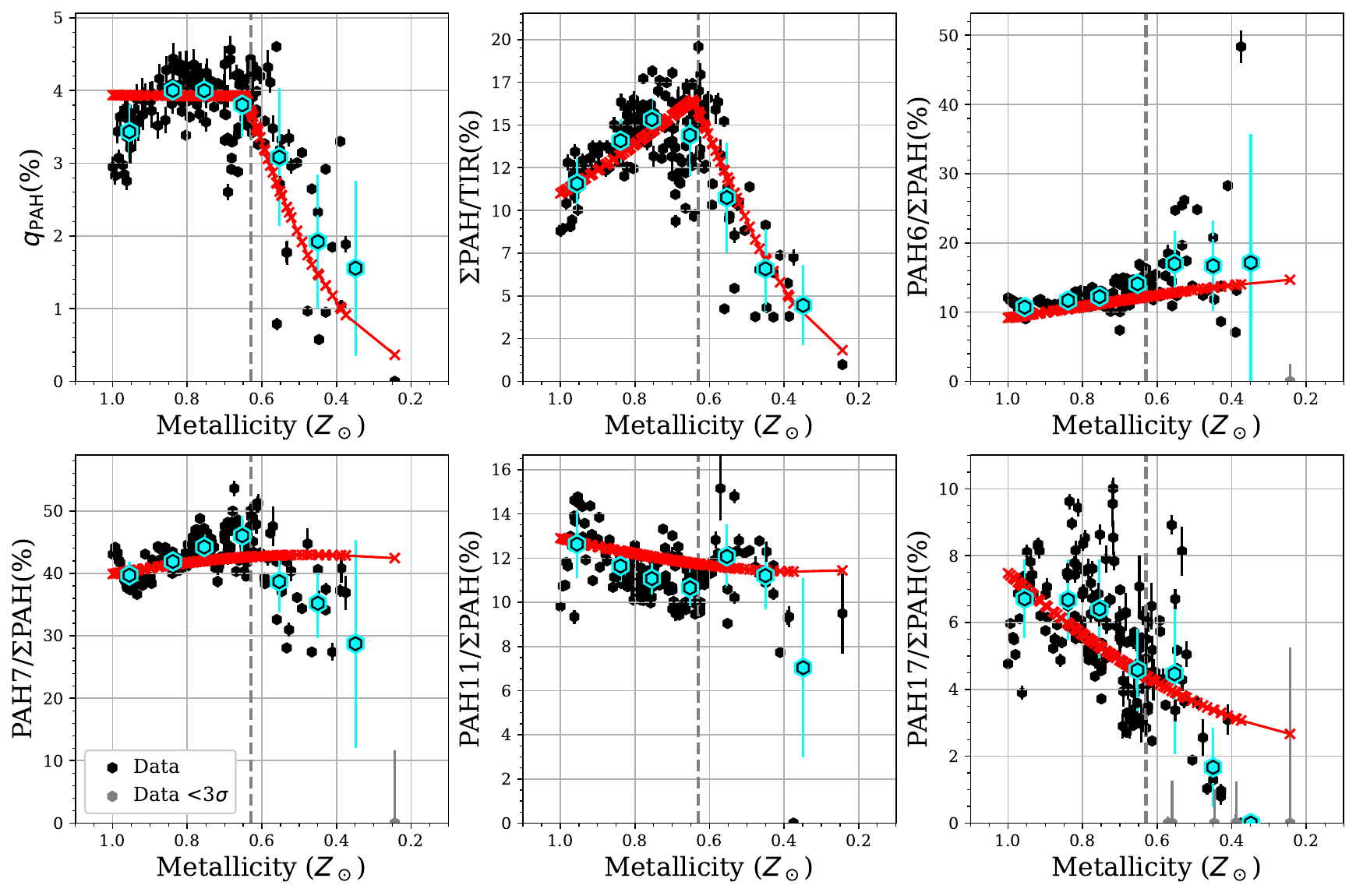}
\caption{\textbf{Inhibited grain growth model summary.} As in Fig.~\ref{figure:guillosumm}. \textbf{Top-Left}: GSDs as a function of metallicity. \textbf{Top-right}: Mapping between average PAH size \aone\ (black) and GSD re-scale factor $B_r$ (pink) and metallicity \textbf{Bottom}: comparisons between six observed PZR trends (black) and our best inhibited growth model (red).}
\label{figure:shreddersumm}
\end{figure*}

\begin{figure}
\includegraphics[width=\linewidth]{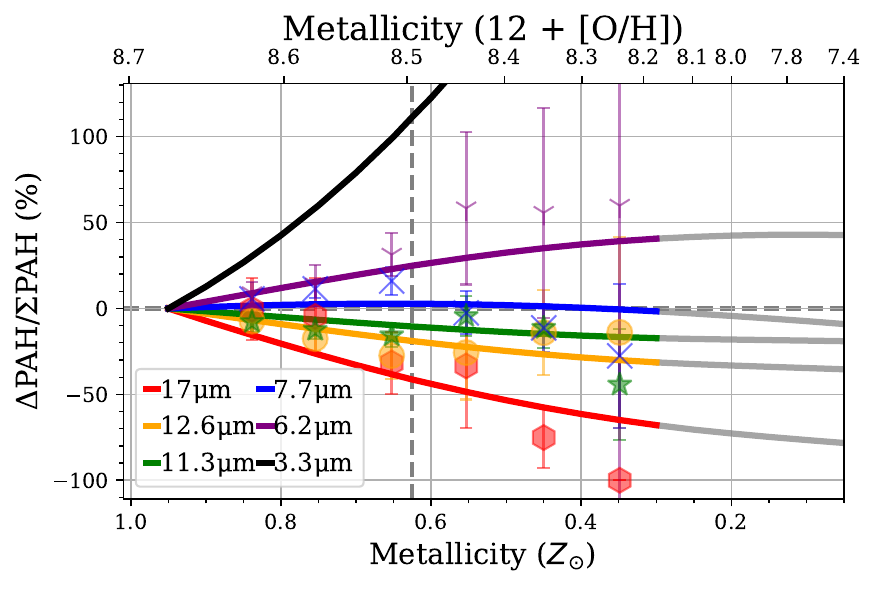}
\caption{
\textbf{Fractional PAH power as a function of metallicity for our best inhibited growth model (\S\ref{sec:shredder})}. The data bins from Figure~\ref{figure:levers} are shown. The gray lines indicate extrapolations of the model to lower metallicities than probed by our observations. Also shown is the model-predicted trend for the fractional change in the relative 3.3\,\micron\ feature brightness.
}
\label{figure:shredlever}
\end{figure}

\subsubsection{Inhibited Growth Model}\label{sec:shredder}

We simulate the effects of inhibited PAH growth by reducing the abundance and fiducial size of PAHs. The fiducial size parameter \aone\ characterizes the location of the first peak in the bi-modal log-normal GSD of carbonaceous dust grains seen in Figure~\ref{figure:sizedist}. To reproduce the observed \qpah, we also use a multiplicative factor $B_r$ that re-scales the entire GSD as metallicity falls.

The functional form of \aone(\zmetal) is assumed to be a power law: \aone~=~5\AA~(\zmetal/\zsolar)$^\beta$, where the value of \aone\ at \zsolar\ is chosen to match the `large' end of sizes considered in \citet{D21}, and $\beta~>~0$. We then explore the parameter space of $\beta\leq6$ in steps of 0.5 by calculating the resulting $\chi^2$ between the the model and 6 primary observed PZR trends (i.e. \qpah, \pahtir, \sixtot, \sevtot, \elevtot, and \teentot). We assume a similar power-law form for $B_r$(\zmetal) but with a different exponent, constrained by minimizing $\chi^2$ between the model and \qpah\ trends.

Figure~\ref{figure:shreddersumm} summarizes the inhibited growth model that most closely matches observations. Since \qpah\ is held constant as metallicity drops from \zsolar\ to \zthresh, the amplitude of the GSD stays approximately constant until metallicity drops below \zthresh. The best value for $\beta$ is 1, with negligible improvements from increasing to 1.5 or 2; future work with more data will allow us to better constrain this parameter.

Since $B_r$ is constrained based on the \qpah-metallicity trend, the functional form of $B_r$ is equivalent to a fit to the \qpah\ trend. Above \zthresh\ $B_r$ increases slightly with metallicity to keep \qpah\ constant while \aone\ shifts to larger sizes. \qpah\ has a small dependence on \aone\ because it is defined as the integral of the GSD for grains with \carbons~$<~10^3$, so as \aone\ is increased, more of the distribution falls above this range and \qpah\ decreases slightly. Below \zthresh\ \qpah\ decreases as a power law with index 2.1 and $B_r$ has this same relation normalized to one at \zthresh.

With \aone\ decreasing linearly as metallicity drops the relative changes in each major PAH feature are reproduced to first order (see Figure~\ref{figure:shreddersumm} and Figure~\ref{figure:shredlever}). Specifically: \teentot\ drops by 50\% at 0.4\,\zsolar\ relative to \zsolar, \elevtot\ drops by $\sim20\%$, \sevtot\ drops by $\sim5\%$, and \sixtot\ rises by $\sim50\%$. There are residual differences between the best model and the observations. For example, it is not possible to match the strong increase in \sixtot\ of almost 75\% between \zsolar\ and 0.4\,\zsolar\ with any value of $\beta$. Similarly we cannot reproduce the $\sim$30\% drop in \sevtot\ after a 10\% rise from \zsolar\ to \zthresh. The model does show this rise and fall in \sevtot, but the amplitude is much smaller. The photo-destruction model results in a better match for this specific trend, but fails significantly for other bands. 

The inhibited growth model does an excellent job reproducing the observed trend between \pahtir\ and metallicity. This trend is \emph{not} explicitly fit; the modeled trend is a result of employing the factor $B_r$ to best match the \qpah\ trend and fitting \aone($Z$) to best match the ratios \rattot. While \qpah\ is constant as metallicity drops from \zsolar\ to \zthresh, \pahtir\ rises because the average PAH size \aone\ decreases, and smaller PAHs emit more per unit mass \citep{D21}. Below \zthresh, \qpah\ begins to drop as well and this causes \pahtir\ to fall faster than the shifting of \aone\ causes it to rise.

\section{Discussion}\label{sec:disc}

\subsection{Inferences from Observational Trends}

As metallicity decreases from \zsolar\ to $\frac{2}{3}$ \zsolar, \pahtir\ remains fairly constant at about 15\% with a scatter of about 2\% (see Figure~\ref{figure:pahtir}). As metallicity decreases below $\frac{2}{3}$\,\zsolar\ \pahtir\ begins to drop following a power law of index 2.6. The slope of this power law is steeper than observed in previous works. \citet{remy15} found a slope of about 1.3 between their $f_{\rm PAH}$ and oxygen abundance for galaxies in the Dwarf Galaxy Survey (DGS) and part of the KINGFISH sample \citep{madden13, KINGFISH}. The difference may be a result of sample variations since \citet{remy15} compares photometric integrated galaxy measurements for the PAH fraction while in this work we study the PAH fraction from spectroscopy within individual galaxies. Future studies in this series with a significantly larger sample will allow us to place better constraints on the power law index and scatter of the \pahtir-metallicity relationship.

Some of the \textsc{Hii} regions are notably offset below the overall \pahtir\ trend with metallicity. Since these regions have a lower \pahtir\ than their metallicity would otherwise imply, this may be evidence of a local PAH luminosity deficit inside ionized gas within \textsc{Hii} regions. Previous works have proposed that the hard radiation environment near young O and B stars destroys small dust grains \citep{povich07, chaste19, egorov23}. In the UV-illuminated neutral and molecular gas of the photo-dissociation region surrounding the \textsc{Hii} region, the small grains survive and have a steady source of photons for excitation. At the $\sim\!300$\,pc scales probed by the apertures used in this work, our H\textsc{ii} region pointings contain a mix of both of these environments, resulting in a moderately lower \pahtir\ in some cases. However, intrinsic features of the PAH spectrum (i.e. band to total and band to band ratios) do not appear to be distinct for \textsc{Hii} and non-\textsc{Hii} regions at a given metallicity, consistent with the mild effects of ISRF intensity and hardness (see Figures~\ref{figure:pahtir} and \ref{figure:hardlever}, respectively). This could indicate that the PAH population remains fairly uniform outside \textsc{Hii} regions, but that there is a deficit of all PAHs relative to larger grains in the ionized gas itself \citep[e.g.][]{sutter24}.

The modeled quantity \qpah\ is typically derived from the brightness at IRAC 8\,\micron\ which only traces the 7.7\,\micron\ PAH feature. In this work we have found that all PAH feature strengths decrease rapidly relative to TIR when metallicity falls below \zthresh, but some features drop faster than others. Specifically, relative to the total dust luminosity, the 6.2\,\micron\ feature decreases more slowly than the 7.7\,\micron\ feature, and the 17\,\micron\ feature decreases significantly faster as metallicity drops, implying that the situation is more complex than an overall decrease in PAH abundance (\qpah). The assumption that the 7.7\,\micron\ feature is representative of all PAH emission underlies many past interpretations based on 8\,\micron\ photometry. The common explanation for the decrease in 8\,\micron\ photometry in such studies is photo-destruction of the smallest PAHs. However, our observations show an increase of 6.2\,\micron/\totpah\ at low metallicity (see Figure~\ref{figure:levers}). The 6.2\,\micron\ feature is more dependent on the presence of small PAHs than 7.7\,\micron, so the increase in 6.2\,\micron/\totpah\ is evidence against this photo-destruction explanation for the PZR trends.

The shift in power between PAH features provides a good diagnostic of what is happening to the PAHs as metallicity changes. We observed a general trend of power shifting from longer wavelength features to shorter wavelength features. This occurs immediately as metallicity decreases below \zsolar, despite the fact that \qpah\ and \pahtir\ remain approximately constant (with bulge ISRF correction applied, see \S\,\ref{sec:bulge}). Previous works have observed similar trends. \citet{jd07} found the ratio \elevteen\ increases from $\sim$1.2 to $\sim$2.2 as metallicity decreases from 0.8\,\zsolar\ to 0.45\,\zsolar.  We find a more significant increase in \elevteen\ over this range, from $\sim$1.5 to $\sim$10.1 (see top panel of Figure~\ref{figure:ratios}). Note the average at 0.45\,\zsolar\ is strongly affected by the points from NGC~2403 while the points from M101 show a lower average of $\sim$2.5 that is more in line with the \citet{jd07} result.

Another nearby low metallicity galaxy with extensive Spitzer-IRS spectroscopy is the SMC \citep[$\sim0.2\,Z_{\odot}$;][]{toribio2017}. Measurements of the PAH fraction and band strengths in the SMC are in good agreement with the trends predicted by our inhibited growth model at that metallicity, and do not match predictions with the photo-destruction or ISRF hardness models.  In particular, \qpah\ for the SMC is $\lesssim$\,1\% \citep{sandstrom2010,chaste19}. The band strengths in the SMC show a similar shift of PAH power between features, namely higher \sixtot\ and \elevtot\ and lower \sevtot\ compared to higher metallicity regions \citep{sandstrom12}. The SMC regions also showed notably low \teentot\ compared to the SINGS average. These trends are all in line with a scenario where inhibited growth leads to smaller PAHs and lower PAH fractions in the low metallicity SMC.

In addition to metallicity, many physical properties vary with radius in spiral galaxies. Of relevance to PAH emission, changes in the radiation environment including the intensity and average photon hardness are expected.  As found in Figure~\ref{figure:Ubar} and \S\,\ref{sec:hardmodel}, these properties are unlikely to drive the observed trends.  Other potentially radially-varying properties we do not model that could affect PAH emission include the phase balance of dense and neutral gas, gross differences in the optical properties of small carbon grains, varying carbon depletion fractions, and the metal content of the gas that exists in the dense phase.

\begin{figure}
\centering
\includegraphics[width=\linewidth]{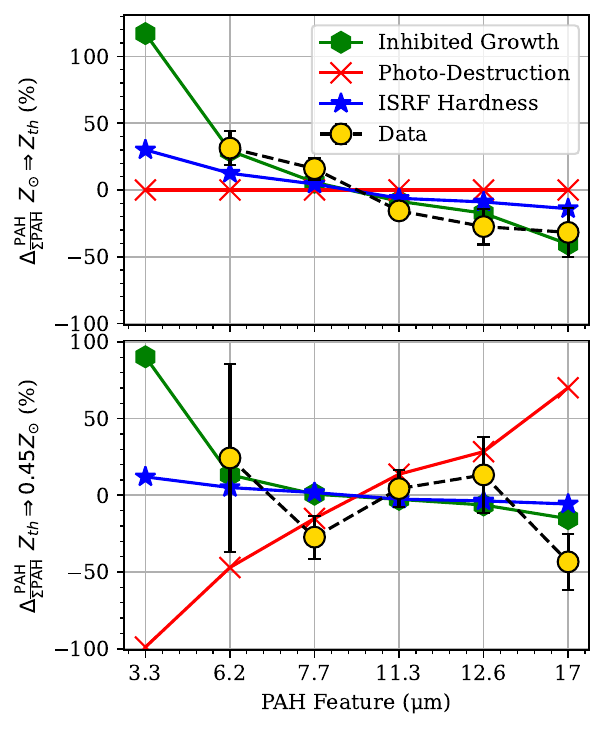}
\caption{\textbf{Summary of observed and modeled \rattot\ differentials}. \textbf{Top}: Percent change in each major PAH feature relative to the sum of all PAH emission, as metallicity decreases from \zsolar\ to \zthresh.  Observations are in black and the three models considered appear as colored curves. 
\textbf{Bottom}: Same as above but for metallicity decreasing from \zthresh\ to 0.45 \zsolar.
Note that the models differ most in their predictions for the 3.3\,\micron\ feature, not included in the current data set.}
\label{figure:summary}
\end{figure}

\vspace{-0.1in}
\subsection{Modeling Interpretation}\label{sec:modeldisc}

We tested three physically-motivated explanations for the observed PZR trends using D21 models: ISRF hardness, photo-destruction, and inhibited grain growth. A summary of how each of these models compares to the observed PAH band ratios is shown in Figure~\ref{figure:summary}. The ISRF hardness model quantified the effect of changing only the incident starlight spectrum with a fixed PAH population. The photo-destruction model tested the most common explanation for the PZR: selective destruction of small grains. The inhibited growth model approximated the effects of reduced accretion of carbon onto grains by decreasing the average size and overall abundance of PAHs.

The ISRF hardness model was intended to simulate a scenario where the PAH GSD remains unchanged and the PZR trends arise purely a result of the increasing radiation hardness as metallicity decreases. This model resulted in the correct band ratio trends, but the magnitude of these trends is too low by a factor of 2--3.  Furthermore, the adopted model explores the \emph{extremes} of such a hardness trend, varying the incident starlight spectrum from that of a 3\,\zsolar\ old bulge down to a \zsolar/20 starburst.  We conclude that starlight hardness alone cannot account for the varying shape of the PAH spectrum, and that the small GSD must also be changing.

If, however, the ratio of far-UV-to-optical absorption cross sections of the PAH grains were substantially increased, then such a model could better match the observations as grain heating rate would exhibit increased sensitivity to the starlight hardness. \citet{donnelly24} also concluded that such a change in the relative absorption cross sections would result in better agreement between the D21 models and PAH emission observed near a low-luminosity active galactic nucleus. Regardless, the radiation hardness model is not capable of reproducing the \qpah\ and \pahtir\ trends as a function of metallicity. As the average exciting photon energy increases, it is not possible for either \qpah\ or \pahtir\ to decrease.  In fact, without changes in the grain population, relative PAH power is expected to \emph{increase} with hardness, as every PAH grain reaches higher stochastic temperatures.

The photo-destruction scenario we modelled in \S\,\ref{sec:guillotine} is commonly invoked to explain the deficit of PAH emission relative to total dust luminosity as a function of decreasing metallicity \citep{madden06, gordon08}. To reproduce the steep \qpah\ decline, the photo-destruction model requires increasing the grain cutoff size $a_\mathrm{min}$ up to $\sim$400\,\carbons\ to match the \qpah\ drop from its \zsolar\ value of $\sim$3.9\% to below 0.5\% at 0.2\,\zsolar. Based on theoretical work, it is highly unlikely for a majority of PAHs smaller than 400\,\carbons\ to be photo-dissociated in ISM at low metallicity. \citet{guha89} found that PAHs smaller than $\sim$\,23\,\carbons\ cannot survive in the ISM under typical solar-neighborhood radiation conditions.
In \citet{guha89} the significantly harder radiation field of a B type star is considered and is found to increase the minimum grain size to $a_\mathrm{min}$\,$\sim$\,35\,\carbons. This is an order of magnitude too low to explain the observed trend in \qpah\ with photo-destruction alone.

Because of this and the strong disagreement in band ratio behavior, we conclude that small grain photo-destruction likely does not play a dominant role in the PZR. We do note that this is based on PAH emission trends at wavelengths longer than 5\,\micron.  It remains possible that a modest amount of photo-destruction may affect the very smallest PAHs traced by the 3.3\,\micron\ feature, which the photo-destruction model predicts will drop precipitously towards low metallicity.

The model that best matches the observed PAH-metallicity trends simulates inhibited grain growth. We implemented this scenario by shifting the average PAH size smaller and reducing the total mass of all carbon grains as metallicity decreases. If individual carbon atoms are accreted and grow on the surfaces of larger grains (bottom-up formation), their reduced availability and longer accretion timescales at lower metallicity could explain the overall shift towards smaller grain sizes, although the increased intensity and hardness of radiation could also play a role in inhibiting growth. 

While the metallicity dependence of the gas-grain accretion timescale supports this interpretation, any physical mechanism which substantially shifts the PAH GSD to smaller sizes would produce similar effects on the PAH spectrum. For instance, grain sizes injected by carbon-rich AGB stars could vary with initial stellar metallicity \citep{otsuka14}, although the nature of any such dependency is not well established. Another possible mechanism is selective processing of the largest PAH grains. However, photo-processing is believed to have the greatest impact on smaller PAHs, and grain erosion from selective sputtering of all large grains would serve to \emph{increase} the fractional PAH luminosity at long wavelengths, which does not match the observations.

Despite the successes of the inhibited growth model, several trends in the observations remain unaccounted for. These may be a result of simplifying assumptions in the D21 models. We have assumed that the radiation field intensity remains fixed at $U=1$. This is the typical value of $U$ extracted from the SED modelling maps with our matched apertures, and there is little effect on the shape or relative strength of the PAH spectrum at such low intensities. On smaller scales (e.g. approaching individual H\textsc{ii} regions), however, this assumption may break down and larger values of $U$ that could lead to multi-photon heating may be present.

Similarly, aside from the radiation hardness model, we have assumed the mMMP radiation field of the solar neighborhood applies to all regions in each galaxy.  We have no direct measure of the shape of the incident SED that illuminates the dust, but the radiation hardness model (\S\,\ref{sec:hardmodel}) shows that even shifting from an M31 bulge SED to a \zsolar/20 10Myr SED cannot reproduce the observed band ratio variations as a function of metallicity. Since our data do not span such a wide range in metallicity, the local radiation fields are unlikely to be changing this rapidly. However, previous works have found evidence that band ratio variations are correlated with radiation field hardness in nearby galaxies \citep{baron24}.

All three models assume unchanging vibrational properties of PAH grain material; if the chemical structure of PAH grains change as metallicity declines -- for example, losing the ability to emit effectively at 17\,\micron -- this could offer an alternative explanation for the success of the inhibited growth model.

Finally, we have also made the assumption that the ionization fraction as a function of grain size remains fixed at the standard value. This was done primarily to reduce the number of free parameters. However, the overall success of our inhibited growth model implies the standard ionization fraction as a function of size is reasonable. Since the ionization fraction is lower for smaller grains and the average grain size in our inhibited growth model decreases, the model naturally results in reduced effective ionization fraction. This is consistent with the observations of the \sevelev\ ratio shown in Figure~\ref{figure:ratios}. 

The strong agreement between the inhibited growth model and the observations emphasizes the role of accretion and regrowth in maintaining the small grain population in the ISM. At low metallicities, these processes become less efficient as the carbon abundance and average size and number of large grains drops \citep{zhukovska08}. This suggests that the majority of small grains are formed by accretion in the denser phases of the ISM and are not directly stellar in origin.

\subsection{Future Work}

The spectral coverage of Spitzer did not include the aromatic feature at 3.3\,\micron. Theoretical and laboratory results show this feature originates almost exclusively from the smallest PAH molecules \citep[$<$~100\,\carbons,][]{marag20, D21}. If these small grains are significantly affected by photo-destruction, we expect that the 3.3\,\micron\ emission will become very weak at sufficiently low metallicity. And yet, the low-metallicity galaxy IIZw40 ($\sim$20\% \zsolar) has been found to have a relatively \emph{high} fractional 3.3\micron\ luminosity \citep[$\sim3\%$,][]{lai20}.

We have also seen little evidence in this work that small PAHs are preferentially destroyed at low metallicity.  In fact, the success of our inhibited growth model suggests that even as the total PAH grain population drops in abundance, smaller PAHs \emph{rise} in relative importance. However, the 3.3\,\micron\ feature is far more sensitive to the abundance of PAHs with \carbons\,$\ll\,100$ than the 6.2 or 7.7\,\micron\ features \citep{hensley23}, so our conclusions in this work cannot rule out photo-destruction thresholds at smaller grain sizes.

If our inhibited grain growth model holds true, we predict the 3.3\,\micron\ fraction should carry a higher fraction of the total emission from PAH bands as metallicity decreases. The best inhibited growth model suggests the 3.3\,\micron\ feature contributes only 2\% of all PAH band emission at solar metallicity, but extrapolating to 0.2\,\zsolar\ we predict 3.3\,\micron\ could contribute about 10\% of all remaining PAH band emission. As shown in Figure~\ref{figure:summary}, the relative behavior of the 3.3\,\micron\ PAH feature has the strongest variation between models. Future studies on the PZR with 3.3\,\micron\ using JWST can disentangle which model or combination of models applies to the smallest PAHs with \carbons\,$\ll\,$100. 

Figure~\ref{figure:pahtir} shows \pahtir\ increases from 10\% to 16\% as metallicity drops from \zsolar\ to \zthresh. We applied a correction for this in \S\,\ref{sec:bulge} before performing the fit under the assumption that the ISRF is softer in the central region of M101. However we also found our inhibited growth model naturally results in an excellent match to the observed \pahtir\ trend as the average grain size shifts larger at higher metallicity. Future work with high metallicity regions dominated by young stars can explore this further and determine if the decrease in \pahtir\ is a result of softer radiation environment or grain growth in high metallicity ISM.  If the trend is truly metallicity dependent, this may suggest that the threshold metallicity $\sim\!\frac{2}{3}$\,\zsolar\ forms a modest local maximum in PAH brightness relative to all other dust.

The high luminosity of PAHs, dropping steeply and changing shape below $\sim\!\frac{2}{3}$\,\zsolar, may indicate that PAH emission could serve as a powerful probe of metal content in galaxies.  JWST has been able to directly detect 3.3\micron\ PAH emission at z\,$\sim$\,4.1 \citep{spilker23}, and  evidence from UV attenuation at z\,$\sim$\,7 may indicate the presence of large quantities of carbonaceous grains at very early times \citep{witstok23}.  A future far-IR observatory with spectroscopic capabilities could offer an effective way to recover dust conditions and metal content in the early Universe \citep{PRIMAGOBOOK}.

\section{Conclusions}\label{sec:concl}

Spitzer-IRS spectroscopic radial strip maps across three nearby spiral galaxies with well-characterized metallicity gradients were used to study the deficit of PAH emission relative to total dust luminosity as metallicity declines. We found \pahtir\ remains relatively constant between \zsolar\ and $\frac{2}{3}$\,\zsolar. Below this threshold, \pahtir\ drops smoothly and rapidly with metallicity. We studied how the PAH emission spectrum changes in strength and form to diagnose the physical origins of the PAH-metallicity relationship. Building on the latest models from \citet{D21}, we reproduced these trends with a model of declining grain size and overall abundance designed to simulate the effects of inhibited PAH growth. The main findings of this work are:

\begin{itemize}
    \item PAH emission constitutes about 15\% of all infrared emission in high metallicity regions of normal star-forming galaxies. Below about $\frac{2}{3}$\,\zsolar, the power emitted in the PAH features begins to drop smoothly with metallicity, approximately as a power law with slope $\sim$2.6.

    \item As its overall luminosity declines with lower metallicity, PAH feature power shifts toward shorter wavelengths, with the 17\micron\ band declining particularly rapidly.  The dominant 7.7\,\micron\ band rises but then drops below the threshold metallicity, potentially due to the competing effects of smaller average grain size and the implicit changes in PAH ionization balance that naturally follow.
    
    \item Models based on the commonly assumed scenario of small grain photo-destruction -- when tuned to match measured trends in \qpah\ -- fail to reproduce observed changes in the PAH spectrum, predicting roughly the \emph{opposite} band ratio trends as observed.  We conclude that photo-destruction cannot be the dominant effect driving the decline in PAH power at low metallicity.
    
    \item Although the harder radiation at low metallicity does shift PAH power to shorter wavelengths, the effects are modest, and modeling suggests the observed trends cannot arise solely from this increased radiation field hardness.  Strong changes to the underlying carbonaceous grain size distribution are needed to reproduce these trends. 
    
    \item A model of inhibited grain growth in which the average size of PAH grains declines while the overall PAH abundance drops can explain most details of the varying PAH spectrum towards low metallicity. This may be evidence of the importance of dust regrowth as grains cycle through dense phases of the ISM.
    
    \item Our best inhibited growth model predicts the 3.3\,\micron\ PAH feature strength relative to the total PAH luminosity will increase from 2\% to 10\% as metallicity decreases from \zsolar\ to 0.2\,\zsolar, although moderate levels of photo-destruction that do not substantially impact longer wavelength bands could limit this value.    
\end{itemize}

\acknowledgements
\noindent CW thanks R. Chandar’s University of Toledo Paper Writing course (2022–2023) for constructive feedback.  We thank K. Gordon, C. Engelbracht, E. Tarantino, D. Narayanan, E. Peeters, L. Hunt, and A. Li for helpful discussions over the years on the PAH-metallicity problem, and D. Berg for insights on metallicity gradients. JDS thanks the Max Plank Institut f\"ur Astronomie for hosting numerous scientific visits which contributed to the development of this work. We gratefully acknowledge support from NASA ADAP grant 80NSSC21K0851. This research has made use of NASA's Astrophysics Data System and Tiny-Tim/Spitzer, developed by John Krist for the Spitzer Science Center. This research was carried out in part at the Jet Propulsion Laboratory, California Institute of Technology, under a contract with NASA.

\bibliographystyle{aasjournal}
\bibliography{references}

\begin{thebibliography}{}
\expandafter\ifx\csname natexlab\endcsname\relax\def\natexlab#1{#1}\fi
\providecommand{\url}[1]{\href{#1}{#1}}
\providecommand{\dodoi}[1]{doi:~\href{http://doi.org/#1}{\nolinkurl{#1}}}
\providecommand{\doeprint}[1]{\href{http://ascl.net/#1}{\nolinkurl{http://ascl.net/#1}}}
\providecommand{\doarXiv}[1]{\href{https://arxiv.org/abs/#1}{\nolinkurl{https://arxiv.org/abs/#1}}}

\bibitem[{{Aniano} {et~al.}(2020){Aniano}, {Draine}, {Hunt}, {Sandstrom},
  {Calzetti}, {Kennicutt}, {Dale}, {Galametz}, {Gordon}, {Leroy}, {Smith},
  {Roussel}, {Sauvage}, {Walter}, {Armus}, {Bolatto}, {Boquien}, {Crocker}, {De
  Looze}, {Donovan Meyer}, {Helou}, {Hinz}, {Johnson}, {Koda}, {Miller},
  {Montiel}, {Murphy}, {Rela{\~n}o}, {Rix}, {Schinnerer}, {Skibba}, {Wolfire},
  \& {Engelbracht}}]{aniano20}
{Aniano}, G., {Draine}, B.~T., {Hunt}, L.~K., {et~al.} 2020, \apj, 889, 150,
  \dodoi{10.3847/1538-4357/ab5fdb}

\bibitem[{{Asplund} {et~al.}(2009){Asplund}, {Grevesse}, {Sauval}, \&
  {Scott}}]{asplund09}
{Asplund}, M., {Grevesse}, N., {Sauval}, A.~J., \& {Scott}, P. 2009, \araa, 47,
  481, \dodoi{10.1146/annurev.astro.46.060407.145222}

\bibitem[{{Baron} {et~al.}(2024){Baron}, {Sandstrom}, {Rosolowsky}, {Egorov},
  {Klessen}, {Leroy}, {Boquien}, {Schinnerer}, {Belfiore}, {Groves},
  {Chastenet}, {Dale}, {Blanc}, {M{\'e}ndez-Delgado}, {Koch}, {Grasha},
  {Chevance}, {Thilker}, {Colombo}, {Williams}, {Pathak}, {Sutter}, {Brown},
  {Wu}, {Peek}, {Emsellem}, {Larson}, \& {Neumann}}]{baron24}
{Baron}, D., {Sandstrom}, K.~M., {Rosolowsky}, E., {et~al.} 2024, arXiv
  e-prints, arXiv:2402.04330, \dodoi{10.48550/arXiv.2402.04330}

\bibitem[{{Bendo} {et~al.}(2012){Bendo}, {Boselli}, {Dariush}, {Pohlen},
  {Roussel}, {Sauvage}, {Smith}, {Wilson}, {Baes}, {Cooray}, {Clements},
  {Cortese}, {Foyle}, {Galametz}, {Gomez}, {Lebouteiller}, {Lu}, {Madden},
  {Mentuch}, {O'Halloran}, {Page}, {Remy}, {Schulz}, \& {Spinoglio}}]{VNGS}
{Bendo}, G.~J., {Boselli}, A., {Dariush}, A., {et~al.} 2012, \mnras, 419, 1833,
  \dodoi{10.1111/j.1365-2966.2011.19735.x}

\bibitem[{{Berg} {et~al.}(2020){Berg}, {Pogge}, {Skillman}, {Croxall},
  {Moustakas}, {Rogers}, \& {Sun}}]{chaosIV}
{Berg}, D.~A., {Pogge}, R.~W., {Skillman}, E.~D., {et~al.} 2020, \apj, 893, 96,
  \dodoi{10.3847/1538-4357/ab7eab}

\bibitem[{{Berg} {et~al.}(2015){Berg}, {Skillman}, {Croxall}, {Pogge},
  {Moustakas}, \& {Johnson-Groh}}]{berg15}
{Berg}, D.~A., {Skillman}, E.~D., {Croxall}, K.~V., {et~al.} 2015, \apj, 806,
  16, \dodoi{10.1088/0004-637X/806/1/16}

\bibitem[{{Berg} {et~al.}(2013){Berg}, {Skillman}, {Garnett}, {Croxall},
  {Marble}, {Smith}, {Gordon}, \& {Kennicutt}}]{berg13}
{Berg}, D.~A., {Skillman}, E.~D., {Garnett}, D.~R., {et~al.} 2013, \apj, 775,
  128, \dodoi{10.1088/0004-637X/775/2/128}

\bibitem[{{Cesarsky} {et~al.}(1996){Cesarsky}, {Lequeux}, {Abergel}, {Perault},
  {Palazzi}, {Madden}, \& {Tran}}]{cesarsky96}
{Cesarsky}, D., {Lequeux}, J., {Abergel}, A., {et~al.} 1996, \aap, 315, L309

\bibitem[{{Chastenet} {et~al.}(2019){Chastenet}, {Sandstrom}, {Chiang},
  {Leroy}, {Utomo}, {Bot}, {Gordon}, {Draine}, {Fukui}, {Onishi}, \&
  {Tsuge}}]{chaste19}
{Chastenet}, J., {Sandstrom}, K., {Chiang}, I.-D., {et~al.} 2019, \apj, 876,
  62, \dodoi{10.3847/1538-4357/ab16cf}

\bibitem[{{Chastenet} {et~al.}(2023){Chastenet}, {Sutter}, {Sandstrom},
  {Belfiore}, {Egorov}, {Larson}, {Leroy}, {Liu}, {Rosolowsky}, {Thilker},
  {Watkins}, {Williams}, {Barnes}, {Bigiel}, {Boquien}, {Chevance}, {Chiang},
  {Dale}, {Kruijssen}, {Emsellem}, {Grasha}, {Groves}, {Hassani}, {Hughes},
  {Kreckel}, {Meidt}, {Rickards Vaught}, {Sardone}, \& {Schinnerer}}]{chaste23}
{Chastenet}, J., {Sutter}, J., {Sandstrom}, K., {et~al.} 2023, \apjl, 944, L11,
  \dodoi{10.3847/2041-8213/acadd7}

\bibitem[{{Choban} {et~al.}(2022){Choban}, {Kere{\v{s}}}, {Hopkins},
  {Sandstrom}, {Hayward}, \& {Faucher-Gigu{\`e}re}}]{choban22}
{Choban}, C.~R., {Kere{\v{s}}}, D., {Hopkins}, P.~F., {et~al.} 2022, \mnras,
  514, 4506, \dodoi{10.1093/mnras/stac1542}

\bibitem[{{Choban} {et~al.}(2024){Choban}, {Kere{\v{s}}}, {Sandstrom},
  {Hopkins}, {Hayward}, \& {Faucher-Gigu{\`e}re}}]{choban24}
{Choban}, C.~R., {Kere{\v{s}}}, D., {Sandstrom}, K.~M., {et~al.} 2024, \mnras,
  529, 2356, \dodoi{10.1093/mnras/stae716}

\bibitem[{{Croxall} {et~al.}(2016){Croxall}, {Pogge}, {Berg}, {Skillman}, \&
  {Moustakas}}]{croxall16}
{Croxall}, K.~V., {Pogge}, R.~W., {Berg}, D.~A., {Skillman}, E.~D., \&
  {Moustakas}, J. 2016, \apj, 830, 4, \dodoi{10.3847/0004-637X/830/1/4}

\bibitem[{{Dale} {et~al.}(2020){Dale}, {Anderson}, {Bran}, {Cox}, {Drake},
  {Lee}, {Pilawa}, {Alexander Slane}, {Soto}, {Jensen}, {Sutter}, {Turner}, \&
  {Kobulnicky}}]{dale20}
{Dale}, D.~A., {Anderson}, K.~R., {Bran}, L.~M., {et~al.} 2020, \aj, 159, 195,
  \dodoi{10.3847/1538-3881/ab7eb2}

\bibitem[{{Donnelly} {et~al.}(2024){Donnelly}, {Smith}, {Draine}, {Togi},
  {Lai}, {Armus}, {Dale}, \& {Charmandaris}}]{donnelly24}
{Donnelly}, G.~P., {Smith}, J.~D.~T., {Draine}, B.~T., {et~al.} 2024, \apj,
  965, 75, \dodoi{10.3847/1538-4357/ad2169}

\bibitem[{{Draine}(2009)}]{draine09}
{Draine}, B.~T. 2009, in Astronomical Society of the Pacific Conference Series,
  Vol. 414, Cosmic Dust - Near and Far, ed. T.~{Henning}, E.~{Gr{\"u}n}, \&
  J.~{Steinacker}, 453

\bibitem[{{Draine} \& {Li}(2007)}]{draineli07}
{Draine}, B.~T., \& {Li}, A. 2007, \apj, 657, 810, \dodoi{10.1086/511055}

\bibitem[{{Draine} {et~al.}(2021){Draine}, {Li}, {Hensley}, {Hunt},
  {Sandstrom}, \& {Smith}}]{D21}
{Draine}, B.~T., {Li}, A., {Hensley}, B.~S., {et~al.} 2021, \apj, 917, 3,
  \dodoi{10.3847/1538-4357/abff51}

\bibitem[{{Draine} {et~al.}(2007){Draine}, {Dale}, {Bendo}, {Gordon}, {Smith},
  {Armus}, {Engelbracht}, {Helou}, {Kennicutt}, {Li}, {Roussel}, {Walter},
  {Calzetti}, {Moustakas}, {Murphy}, {Rieke}, {Bot}, {Hollenbach}, {Sheth}, \&
  {Teplitz}}]{draine07sings}
{Draine}, B.~T., {Dale}, D.~A., {Bendo}, G., {et~al.} 2007, \apj, 663, 866,
  \dodoi{10.1086/518306}

\bibitem[{{Draine} {et~al.}(2014){Draine}, {Aniano}, {Krause}, {Groves},
  {Sandstrom}, {Braun}, {Leroy}, {Klaas}, {Linz}, {Rix}, {Schinnerer},
  {Schmiedeke}, \& {Walter}}]{andromeda}
{Draine}, B.~T., {Aniano}, G., {Krause}, O., {et~al.} 2014, \apj, 780, 172,
  \dodoi{10.1088/0004-637X/780/2/172}

\bibitem[{{Duley}(2009)}]{duley09}
{Duley}, W.~W. 2009, \apj, 705, 446, \dodoi{10.1088/0004-637X/705/1/446}

\bibitem[{{Dwek} \& {Scalo}(1980)}]{dwek80}
{Dwek}, E., \& {Scalo}, J.~M. 1980, \apj, 239, 193, \dodoi{10.1086/158100}

\bibitem[{{Egorov} {et~al.}(2023){Egorov}, {Kreckel}, {Sandstrom}, {Leroy},
  {Glover}, {Groves}, {Kruijssen}, {Barnes}, {Belfiore}, {Bigiel}, {Blanc},
  {Boquien}, {Cao}, {Chastenet}, {Chevance}, {Congiu}, {Dale}, {Emsellem},
  {Grasha}, {Klessen}, {Larson}, {Liu}, {Murphy}, {Pan}, {Pessa}, {Pety},
  {Rosolowsky}, {Scheuermann}, {Schinnerer}, {Sutter}, {Thilker}, {Watkins}, \&
  {Williams}}]{egorov23}
{Egorov}, O.~V., {Kreckel}, K., {Sandstrom}, K.~M., {et~al.} 2023, \apjl, 944,
  L16, \dodoi{10.3847/2041-8213/acac92}

\bibitem[{{Eldridge} {et~al.}(2017){Eldridge}, {Stanway}, {Xiao}, {McClelland},
  {Taylor}, {Ng}, {Greis}, \& {Bray}}]{bpass}
{Eldridge}, J.~J., {Stanway}, E.~R., {Xiao}, L., {et~al.} 2017, \pasa, 34,
  e058, \dodoi{10.1017/pasa.2017.51}

\bibitem[{{Engelbracht} {et~al.}(2005){Engelbracht}, {Gordon}, {Rieke},
  {Werner}, {Dale}, \& {Latter}}]{engel05}
{Engelbracht}, C.~W., {Gordon}, K.~D., {Rieke}, G.~H., {et~al.} 2005, \apjl,
  628, L29, \dodoi{10.1086/432613}

\bibitem[{{Engelbracht} {et~al.}(2008){Engelbracht}, {Rieke}, {Gordon},
  {Smith}, {Werner}, {Moustakas}, {Willmer}, \& {Vanzi}}]{engel08}
{Engelbracht}, C.~W., {Rieke}, G.~H., {Gordon}, K.~D., {et~al.} 2008, \apj,
  678, 804, \dodoi{10.1086/529513}

\bibitem[{{Fisher} \& {Drory}(2010)}]{fisher10}
{Fisher}, D.~B., \& {Drory}, N. 2010, \apj, 716, 942,
  \dodoi{10.1088/0004-637X/716/2/942}

\bibitem[{{Galametz} {et~al.}(2013){Galametz}, {Kennicutt}, {Calzetti},
  {Aniano}, {Draine}, {Boquien}, {Brandl}, {Croxall}, {Dale}, {Engelbracht},
  {Gordon}, {Groves}, {Hao}, {Helou}, {Hinz}, {Hunt}, {Johnson}, {Li},
  {Murphy}, {Roussel}, {Sandstrom}, {Skibba}, \& {Tabatabaei}}]{galam13}
{Galametz}, M., {Kennicutt}, R.~C., {Calzetti}, D., {et~al.} 2013, \mnras, 431,
  1956, \dodoi{10.1093/mnras/stt313}

\bibitem[{{Galliano} {et~al.}(2008){Galliano}, {Dwek}, \&
  {Chanial}}]{galliano08}
{Galliano}, F., {Dwek}, E., \& {Chanial}, P. 2008, \apj, 672, 214,
  \dodoi{10.1086/523621}

\bibitem[{{Galliano} {et~al.}(2018){Galliano}, {Galametz}, \&
  {Jones}}]{galliano18}
{Galliano}, F., {Galametz}, M., \& {Jones}, A.~P. 2018, \araa, 56, 673,
  \dodoi{10.1146/annurev-astro-081817-051900}

\bibitem[{{Garc{\'\i}a-Benito} {et~al.}(2017){Garc{\'\i}a-Benito},
  {Gonz{\'a}lez Delgado}, {P{\'e}rez}, {Cid Fernandes}, {Cortijo-Ferrero},
  {L{\'o}pez Fern{\'a}ndez}, {de Amorim}, {Lacerda}, {Vale Asari}, \&
  {S{\'a}nchez}}]{garcia17}
{Garc{\'\i}a-Benito}, R., {Gonz{\'a}lez Delgado}, R.~M., {P{\'e}rez}, E.,
  {et~al.} 2017, \aap, 608, A27, \dodoi{10.1051/0004-6361/201731357}

\bibitem[{{Gordon} {et~al.}(2008){Gordon}, {Engelbracht}, {Rieke}, {Misselt},
  {Smith}, \& {Kennicutt}}]{gordon08}
{Gordon}, K.~D., {Engelbracht}, C.~W., {Rieke}, G.~H., {et~al.} 2008, \apj,
  682, 336, \dodoi{10.1086/589567}

\bibitem[{{Groves} {et~al.}(2012){Groves}, {Krause}, {Sandstrom}, {Schmiedeke},
  {Leroy}, {Linz}, {Kapala}, {Rix}, {Schinnerer}, {Tabatabaei}, {Walter}, \&
  {da Cunha}}]{groves12}
{Groves}, B., {Krause}, O., {Sandstrom}, K., {et~al.} 2012, \mnras, 426, 892,
  \dodoi{10.1111/j.1365-2966.2012.21696.x}

\bibitem[{{Guhathakurta} \& {Draine}(1989)}]{guha89}
{Guhathakurta}, P., \& {Draine}, B.~T. 1989, \apj, 345, 230,
  \dodoi{10.1086/167899}

\bibitem[{{Haynes} {et~al.}(2010){Haynes}, {Cannon}, {Skillman}, {Jackson}, \&
  {Gehrz}}]{haynes10}
{Haynes}, K., {Cannon}, J.~M., {Skillman}, E.~D., {Jackson}, D.~C., \& {Gehrz},
  R. 2010, \apj, 724, 215, \dodoi{10.1088/0004-637X/724/1/215}

\bibitem[{{Hensley} \& {Draine}(2023)}]{hensley23}
{Hensley}, B.~S., \& {Draine}, B.~T. 2023, \apj, 948, 55,
  \dodoi{10.3847/1538-4357/acc4c2}

\bibitem[{{Hunt} {et~al.}(2010){Hunt}, {Thuan}, {Izotov}, \&
  {Sauvage}}]{hunt10}
{Hunt}, L.~K., {Thuan}, T.~X., {Izotov}, Y.~I., \& {Sauvage}, M. 2010, \apj,
  712, 164, \dodoi{10.1088/0004-637X/712/1/164}

\bibitem[{{Jackson} {et~al.}(2006){Jackson}, {Cannon}, {Skillman}, {Lee},
  {Gehrz}, {Woodward}, \& {Polomski}}]{jackson06}
{Jackson}, D.~C., {Cannon}, J.~M., {Skillman}, E.~D., {et~al.} 2006, \apj, 646,
  192, \dodoi{10.1086/504707}

\bibitem[{{Kassis} {et~al.}(2006){Kassis}, {Adams}, {Campbell}, {Deutsch},
  {Hora}, {Jackson}, \& {Tollestrup}}]{kassis06}
{Kassis}, M., {Adams}, J.~D., {Campbell}, M.~F., {et~al.} 2006, \apj, 637, 823,
  \dodoi{10.1086/498404}

\bibitem[{{Kennicutt} {et~al.}(2003){Kennicutt}, {Bresolin}, \&
  {Garnett}}]{kenn03}
{Kennicutt}, Robert~C., J., {Bresolin}, F., \& {Garnett}, D.~R. 2003, \apj,
  591, 801, \dodoi{10.1086/375398}

\bibitem[{{Kennicutt} {et~al.}(2011){Kennicutt}, {Calzetti}, {Aniano},
  {Appleton}, {Armus}, {Beir{\~a}o}, {Bolatto}, {Brandl}, {Crocker}, {Croxall},
  {Dale}, {Donovan Meyer}, {Draine}, {Engelbracht}, {Galametz}, {Gordon},
  {Groves}, {Hao}, {Helou}, {Hinz}, {Hunt}, {Johnson}, {Koda}, {Krause},
  {Leroy}, {Li}, {Meidt}, {Montiel}, {Murphy}, {Rahman}, {Rix}, {Roussel},
  {Sandstrom}, {Sauvage}, {Schinnerer}, {Skibba}, {Smith}, {Srinivasan},
  {Vigroux}, {Walter}, {Wilson}, {Wolfire}, \& {Zibetti}}]{KINGFISH}
{Kennicutt}, R.~C., {Calzetti}, D., {Aniano}, G., {et~al.} 2011, \pasp, 123,
  1347, \dodoi{10.1086/663818}

\bibitem[{{Khramtsova} {et~al.}(2014){Khramtsova}, {Wiebe}, {Lozinskaya}, \&
  {Egorov}}]{khramtsova14}
{Khramtsova}, M.~S., {Wiebe}, D.~S., {Lozinskaya}, T.~A., \& {Egorov}, O.~V.
  2014, \mnras, 444, 757, \dodoi{10.1093/mnras/stu1482}

\bibitem[{{Kreckel} {et~al.}(2019){Kreckel}, {Ho}, {Blanc}, {Groves},
  {Santoro}, {Schinnerer}, {Bigiel}, {Chevance}, {Congiu}, {Emsellem}, {Faesi},
  {Glover}, {Grasha}, {Kruijssen}, {Lang}, {Leroy}, {Meidt}, {McElroy}, {Pety},
  {Rosolowsky}, {Saito}, {Sandstrom}, {Sanchez-Blazquez}, \&
  {Schruba}}]{kreckel19}
{Kreckel}, K., {Ho}, I.~T., {Blanc}, G.~A., {et~al.} 2019, \apj, 887, 80,
  \dodoi{10.3847/1538-4357/ab5115}

\bibitem[{{Lai} {et~al.}(2020){Lai}, {Smith}, {Baba}, {Spoon}, \&
  {Imanishi}}]{lai20}
{Lai}, T. S.~Y., {Smith}, J.~D.~T., {Baba}, S., {Spoon}, H. W.~W., \&
  {Imanishi}, M. 2020, \apj, 905, 55, \dodoi{10.3847/1538-4357/abc002}

\bibitem[{{Le Page} {et~al.}(2003){Le Page}, {Snow}, \& {Bierbaum}}]{lepage03}
{Le Page}, V., {Snow}, T.~P., \& {Bierbaum}, V.~M. 2003, \apj, 584, 316,
  \dodoi{10.1086/345595}

\bibitem[{{Lebouteiller} {et~al.}(2011){Lebouteiller}, {Bernard-Salas},
  {Whelan}, {Brandl}, {Galliano}, {Charmandaris}, {Madden}, \&
  {Kunth}}]{lebouteiller11}
{Lebouteiller}, V., {Bernard-Salas}, J., {Whelan}, D.~G., {et~al.} 2011, \apj,
  728, 45, \dodoi{10.1088/0004-637X/728/1/45}

\bibitem[{{Li}(2020)}]{li20}
{Li}, A. 2020, Nature Astronomy, 4, 339, \dodoi{10.1038/s41550-020-1051-1}

\bibitem[{{Li} \& {Draine}(2001)}]{lidraine01}
{Li}, A., \& {Draine}, B.~T. 2001, \apj, 554, 778, \dodoi{10.1086/323147}

\bibitem[{{Madden} {et~al.}(2006){Madden}, {Galliano}, {Jones}, \&
  {Sauvage}}]{madden06}
{Madden}, S.~C., {Galliano}, F., {Jones}, A.~P., \& {Sauvage}, M. 2006, \aap,
  446, 877, \dodoi{10.1051/0004-6361:20053890}

\bibitem[{{Madden} {et~al.}(2013){Madden}, {R{\'e}my-Ruyer}, {Galametz},
  {Cormier}, {Lebouteiller}, {Galliano}, {Hony}, {Bendo}, {Smith}, {Pohlen},
  {Roussel}, {Sauvage}, {Wu}, {Sturm}, {Poglitsch}, {Contursi}, {Doublier},
  {Baes}, {Barlow}, {Boselli}, {Boquien}, {Carlson}, {Ciesla}, {Cooray},
  {Cortese}, {de Looze}, {Irwin}, {Isaak}, {Kamenetzky}, {Karczewski}, {Lu},
  {MacHattie}, {O'Halloran}, {Parkin}, {Rangwala}, {Schirm}, {Schulz},
  {Spinoglio}, {Vaccari}, {Wilson}, \& {Wozniak}}]{madden13}
{Madden}, S.~C., {R{\'e}my-Ruyer}, A., {Galametz}, M., {et~al.} 2013, \pasp,
  125, 600, \dodoi{10.1086/671138}

\bibitem[{{Maragkoudakis} {et~al.}(2020){Maragkoudakis}, {Peeters}, \&
  {Ricca}}]{marag20}
{Maragkoudakis}, A., {Peeters}, E., \& {Ricca}, A. 2020, \mnras, 494, 642,
  \dodoi{10.1093/mnras/staa681}

\bibitem[{{Massey} {et~al.}(2005){Massey}, {Puls}, {Pauldrach}, {Bresolin},
  {Kudritzki}, \& {Simon}}]{massey05}
{Massey}, P., {Puls}, J., {Pauldrach}, A.~W.~A., {et~al.} 2005, \apj, 627, 477,
  \dodoi{10.1086/430417}

\bibitem[{{Mathis} {et~al.}(1983){Mathis}, {Mezger}, \& {Panagia}}]{mmp83}
{Mathis}, J.~S., {Mezger}, P.~G., \& {Panagia}, N. 1983, \aap, 128, 212

\bibitem[{{Micelotta} {et~al.}(2010{\natexlab{a}}){Micelotta}, {Jones}, \&
  {Tielens}}]{micel10a}
{Micelotta}, E.~R., {Jones}, A.~P., \& {Tielens}, A.~G.~G.~M.
  2010{\natexlab{a}}, \aap, 510, A36, \dodoi{10.1051/0004-6361/200911682}

\bibitem[{{Micelotta} {et~al.}(2010{\natexlab{b}}){Micelotta}, {Jones}, \&
  {Tielens}}]{micel10b}
---. 2010{\natexlab{b}}, \aap, 510, A37, \dodoi{10.1051/0004-6361/200911683}

\bibitem[{{Moullet} {et~al.}(2023){Moullet}, {Kataria}, {Lis}, {Unwin},
  {Hasegawa}, {Mills}, {Battersby}, {Roc}, \& {Meixner}}]{PRIMAGOBOOK}
{Moullet}, A., {Kataria}, T., {Lis}, D., {et~al.} 2023, arXiv e-prints,
  arXiv:2310.20572, \dodoi{10.48550/arXiv.2310.20572}

\bibitem[{{Mu{\~n}oz-Mateos} {et~al.}(2009){Mu{\~n}oz-Mateos}, {Gil de Paz},
  {Boissier}, {Zamorano}, {Dale}, {P{\'e}rez-Gonz{\'a}lez}, {Gallego},
  {Madore}, {Bendo}, {Thornley}, {Draine}, {Boselli}, {Buat}, {Calzetti},
  {Moustakas}, \& {Kennicutt}}]{munoz09}
{Mu{\~n}oz-Mateos}, J.~C., {Gil de Paz}, A., {Boissier}, S., {et~al.} 2009,
  \apj, 701, 1965, \dodoi{10.1088/0004-637X/701/2/1965}

\bibitem[{{Narayanan} {et~al.}(2023){Narayanan}, {Smith}, {Hensley}, {Li},
  {Hu}, {Sandstrom}, {Torrey}, {Vogelsberger}, {Marinacci}, \&
  {Sales}}]{desika23}
{Narayanan}, D., {Smith}, J. D.~T., {Hensley}, B.~S., {et~al.} 2023, \apj, 951,
  100, \dodoi{10.3847/1538-4357/accf8d}

\bibitem[{{O'Halloran} {et~al.}(2006){O'Halloran}, {Satyapal}, \&
  {Dudik}}]{ohall06}
{O'Halloran}, B., {Satyapal}, S., \& {Dudik}, R.~P. 2006, \apj, 641, 795,
  \dodoi{10.1086/500529}

\bibitem[{{Otsuka} {et~al.}(2014){Otsuka}, {Kemper}, {Cami}, {Peeters}, \&
  {Bernard-Salas}}]{otsuka14}
{Otsuka}, M., {Kemper}, F., {Cami}, J., {Peeters}, E., \& {Bernard-Salas}, J.
  2014, \mnras, 437, 2577, \dodoi{10.1093/mnras/stt2070}

\bibitem[{{Povich} {et~al.}(2007){Povich}, {Stone}, {Churchwell}, {Zweibel},
  {Wolfire}, {Babler}, {Indebetouw}, {Meade}, \& {Whitney}}]{povich07}
{Povich}, M.~S., {Stone}, J.~M., {Churchwell}, E., {et~al.} 2007, \apj, 660,
  346, \dodoi{10.1086/513073}

\bibitem[{{R{\'e}my-Ruyer} {et~al.}(2014){R{\'e}my-Ruyer}, {Madden},
  {Galliano}, {Galametz}, {Takeuchi}, {Asano}, {Zhukovska}, {Lebouteiller},
  {Cormier}, {Jones}, {Bocchio}, {Baes}, {Bendo}, {Boquien}, {Boselli},
  {DeLooze}, {Doublier-Pritchard}, {Hughes}, {Karczewski}, \&
  {Spinoglio}}]{remy14}
{R{\'e}my-Ruyer}, A., {Madden}, S.~C., {Galliano}, F., {et~al.} 2014, \aap,
  563, A31, \dodoi{10.1051/0004-6361/201322803}

\bibitem[{{R{\'e}my-Ruyer} {et~al.}(2015){R{\'e}my-Ruyer}, {Madden},
  {Galliano}, {Lebouteiller}, {Baes}, {Bendo}, {Boselli}, {Ciesla}, {Cormier},
  {Cooray}, {Cortese}, {De Looze}, {Doublier-Pritchard}, {Galametz}, {Jones},
  {Karczewski}, {Lu}, \& {Spinoglio}}]{remy15}
---. 2015, \aap, 582, A121, \dodoi{10.1051/0004-6361/201526067}

\bibitem[{{Rigopoulou} {et~al.}(2021){Rigopoulou}, {Barale}, {Clary}, {Shan},
  {Alonso-Herrero}, {Garc{\'\i}a-Bernete}, {Hunt}, {Kerkeni},
  {Pereira-Santaella}, \& {Roche}}]{rigopoulou21}
{Rigopoulou}, D., {Barale}, M., {Clary}, D.~C., {et~al.} 2021, \mnras, 504,
  5287, \dodoi{10.1093/mnras/stab959}

\bibitem[{{Rogers} {et~al.}(2021){Rogers}, {Skillman}, {Pogge}, {Berg},
  {Moustakas}, {Croxall}, \& {Sun}}]{chaosVI}
{Rogers}, N. S.~J., {Skillman}, E.~D., {Pogge}, R.~W., {et~al.} 2021, \apj,
  915, 21, \dodoi{10.3847/1538-4357/abf8b9}

\bibitem[{{Sandstrom} {et~al.}(2010){Sandstrom}, {Bolatto}, {Draine}, {Bot}, \&
  {Stanimirovi{\'c}}}]{sandstrom2010}
{Sandstrom}, K.~M., {Bolatto}, A.~D., {Draine}, B.~T., {Bot}, C., \&
  {Stanimirovi{\'c}}, S. 2010, \apj, 715, 701,
  \dodoi{10.1088/0004-637X/715/2/701}

\bibitem[{{Sandstrom} {et~al.}(2009){Sandstrom}, {Bolatto}, {Stanimirovi{\'c}},
  {van Loon}, \& {Smith}}]{sandstrom09}
{Sandstrom}, K.~M., {Bolatto}, A.~D., {Stanimirovi{\'c}}, S., {van Loon},
  J.~T., \& {Smith}, J.~D.~T. 2009, \apj, 696, 2138,
  \dodoi{10.1088/0004-637X/696/2/2138}

\bibitem[{{Sandstrom} {et~al.}(2012){Sandstrom}, {Bolatto}, {Bot}, {Draine},
  {Ingalls}, {Israel}, {Jackson}, {Leroy}, {Li}, {Rubio}, {Simon}, {Smith},
  {Stanimirovi{\'c}}, {Tielens}, \& {van Loon}}]{sandstrom12}
{Sandstrom}, K.~M., {Bolatto}, A.~D., {Bot}, C., {et~al.} 2012, \apj, 744, 20,
  \dodoi{10.1088/0004-637X/744/1/20}

\bibitem[{{Shim} {et~al.}(2023){Shim}, {Hwang}, {Jeong}, {Toba}, {Kim}, {Kim},
  {Song}, {Hashimoto}, {Nakagawa}, {Nanni}, {Pearson}, \& {Takagi}}]{shim23}
{Shim}, H., {Hwang}, H.~S., {Jeong}, W.-S., {et~al.} 2023, \aj, 165, 31,
  \dodoi{10.3847/1538-3881/aca09c}

\bibitem[{{Shivaei} {et~al.}(2017){Shivaei}, {Reddy}, {Shapley}, {Siana},
  {Kriek}, {Mobasher}, {Coil}, {Freeman}, {Sanders}, {Price}, {Azadi}, \&
  {Zick}}]{shivaei17}
{Shivaei}, I., {Reddy}, N.~A., {Shapley}, A.~E., {et~al.} 2017, \apj, 837, 157,
  \dodoi{10.3847/1538-4357/aa619c}

\bibitem[{{Shivaei} {et~al.}(2024){Shivaei}, {Alberts}, {Florian}, {Rieke},
  {Wuyts}, {Bodansky}, {Bunker}, {Cameron}, {Curti}, {D'Eugenio},
  {Dudzeviciute}, {Kramarenko}, {Ji}, {Johnson}, {Lyu}, {Matthee}, {Morrison},
  {Naidu}, {Reddy}, {Robertson}, {P{\'e}rez-Gonz{\'a}lez}, {Sun}, {Tacchella},
  {Whitaker}, {Williams}, {Willmer}, {Witstok}, {Xiao}, \& {Zhu}}]{shivaei24}
{Shivaei}, I., {Alberts}, S., {Florian}, M., {et~al.} 2024, arXiv e-prints,
  arXiv:2402.07989, \dodoi{10.48550/arXiv.2402.07989}

\bibitem[{{Skillman} {et~al.}(2020){Skillman}, {Berg}, {Pogge}, {Moustakas},
  {Rogers}, \& {Croxall}}]{chaosV}
{Skillman}, E.~D., {Berg}, D.~A., {Pogge}, R.~W., {et~al.} 2020, \apj, 894,
  138, \dodoi{10.3847/1538-4357/ab86ae}

\bibitem[{{Smith} {et~al.}(2007){Smith}, {Draine}, {Dale}, {Moustakas},
  {Kennicutt}, {Helou}, {Armus}, {Roussel}, {Sheth}, {Bendo}, {Buckalew},
  {Calzetti}, {Engelbracht}, {Gordon}, {Hollenbach}, {Li}, {Malhotra},
  {Murphy}, \& {Walter}}]{jd07}
{Smith}, J.~D.~T., {Draine}, B.~T., {Dale}, D.~A., {et~al.} 2007, \apj, 656,
  770, \dodoi{10.1086/510549}

\bibitem[{{Spilker} {et~al.}(2023){Spilker}, {Phadke}, {Aravena}, {Archipley},
  {Bayliss}, {Birkin}, {B{\'e}thermin}, {Burgoyne}, {Cathey}, {Chapman},
  {Dahle}, {Gonzalez}, {Gururajan}, {Hayward}, {Hezaveh}, {Hill}, {Hutchison},
  {Kim}, {Kim}, {Law}, {Legin}, {Malkan}, {Marrone}, {Murphy}, {Narayanan},
  {Navarre}, {Olivier}, {Rich}, {Rigby}, {Reuter}, {Rhoads}, {Sharon}, {Smith},
  {Solimano}, {Sulzenauer}, {Vieira}, {Vizgan}, {Wei{\ss}}, \&
  {Whitaker}}]{spilker23}
{Spilker}, J.~S., {Phadke}, K.~A., {Aravena}, M., {et~al.} 2023, \nat, 618,
  708, \dodoi{10.1038/s41586-023-05998-6}

\bibitem[{{Sutter} {et~al.}(2024){Sutter}, {Sandstrom}, {Chastenet}, {Leroy},
  {Koch}, {Williams}, {Chown}, {Belfiore}, {Bigiel}, {Boquien}, {Cao},
  {Chevance}, {Dale}, {Egorov}, {Glover}, {Groves}, {Klessen}, {Kreckel},
  {Larson}, {Oakes}, {Pathak}, {Ramambason}, {Rosolowsky}, \&
  {Watkins}}]{sutter24}
{Sutter}, J., {Sandstrom}, K., {Chastenet}, J., {et~al.} 2024, arXiv e-prints,
  arXiv:2405.15102, \dodoi{10.48550/arXiv.2405.15102}

\bibitem[{{Tielens}(2008)}]{tiel08}
{Tielens}, A.~G.~G.~M. 2008, \araa, 46, 289,
  \dodoi{10.1146/annurev.astro.46.060407.145211}

\bibitem[{{Toribio San Cipriano} {et~al.}(2017){Toribio San Cipriano},
  {Dom{\'\i}nguez-Guzm{\'a}n}, {Esteban}, {Garc{\'\i}a-Rojas}, {Mesa-Delgado},
  {Bresolin}, {Rodr{\'\i}guez}, \& {Sim{\'o}n-D{\'\i}az}}]{toribio2017}
{Toribio San Cipriano}, L., {Dom{\'\i}nguez-Guzm{\'a}n}, G., {Esteban}, C.,
  {et~al.} 2017, \mnras, 467, 3759, \dodoi{10.1093/mnras/stx328}

\bibitem[{{Whitcomb} {et~al.}(2020){Whitcomb}, {Sandstrom}, {Murphy}, \&
  {Linden}}]{whitcomb20}
{Whitcomb}, C.~M., {Sandstrom}, K., {Murphy}, E.~J., \& {Linden}, S. 2020,
  \apj, 901, 47, \dodoi{10.3847/1538-4357/abaef6}

\bibitem[{{Witstok} {et~al.}(2023){Witstok}, {Shivaei}, {Smit}, {Maiolino},
  {Carniani}, {Curtis-Lake}, {Ferruit}, {Arribas}, {Bunker}, {Cameron},
  {Charlot}, {Chevallard}, {Curti}, {de Graaff}, {D'Eugenio}, {Giardino},
  {Looser}, {Rawle}, {Rodr{\'\i}guez del Pino}, {Willott}, {Alberts}, {Baker},
  {Boyett}, {Egami}, {Eisenstein}, {Endsley}, {Hainline}, {Ji}, {Johnson},
  {Kumari}, {Lyu}, {Nelson}, {Perna}, {Rieke}, {Robertson}, {Sandles},
  {Saxena}, {Scholtz}, {Sun}, {Tacchella}, {Williams}, \&
  {Willmer}}]{witstok23}
{Witstok}, J., {Shivaei}, I., {Smit}, R., {et~al.} 2023, \nat, 621, 267,
  \dodoi{10.1038/s41586-023-06413-w}

\bibitem[{{Zang} {et~al.}(2022){Zang}, {Maragkoudakis}, \& {Peeters}}]{zang22}
{Zang}, R.~X., {Maragkoudakis}, A., \& {Peeters}, E. 2022, \mnras, 511, 5142,
  \dodoi{10.1093/mnras/stac214}

\bibitem[{{Zhukovska} {et~al.}(2008){Zhukovska}, {Gail}, \&
  {Trieloff}}]{zhukovska08}
{Zhukovska}, S., {Gail}, H.~P., \& {Trieloff}, M. 2008, \aap, 479, 453,
  \dodoi{10.1051/0004-6361:20077789}

\bibitem[{{Zhukovska} \& {Henning}(2013)}]{zhukovska13}
{Zhukovska}, S., \& {Henning}, T. 2013, \aap, 555, A99,
  \dodoi{10.1051/0004-6361/201321368}

\end{thebibliography}

\end{document}